\documentclass[aps,prx,notitlepage,superscriptaddress,longbibliography,twocolumn,nofootinbib,floatfix]{revtex4-2}

\usepackage[utf8]{inputenc}
\usepackage{graphicx}
\usepackage{hyperref}
\usepackage[dvipsnames]{xcolor}

\usepackage{tikz,tikz-3dplot}
\usetikzlibrary{decorations.pathreplacing}
\usepackage{pgfplots}

\usepackage[compat=1.0.0]{tikz-feynman}
\usepackage{simpler-wick}

\usepackage{lipsum}

\usepackage{amsmath,amsthm,amssymb,mathtools}
\usepackage{multirow}
\usepackage{braket}
\usepackage{bbm,bm}
\usepackage[bb=boondox]{mathalfa}
\usepackage{dsfont}

\usepackage{iftex}
\usetikzlibrary{quantikz}

\DeclareFontFamily{U}{matha}{\hyphenchar\font45}
\DeclareFontShape{U}{matha}{m}{n}{
      <5> <6> <7> <8> <9> <10> gen * matha
      <10.95> matha10 <12> <14.4> <17.28> <20.74> <24.88> matha12
      }{}
\DeclareSymbolFont{matha}{U}{matha}{m}{n}
\DeclareMathSymbol{\muparrow}{3}{matha}{"D2}
\DeclareMathSymbol{\mdownarrow}{3}{matha}{"D3}
\DeclareMathSymbol{\mupdownarrow}{3}{matha}{"D9}

\DeclareFontFamily{U}{mathb}{\hyphenchar\font45}
\DeclareFontShape{U}{mathb}{m}{n}{
      <5> <6> <7> <8> <9> <10> gen * mathb
      <10.95> mathb10 <12> <14.4> <17.28> <20.74> <24.88> mathb12
      }{}
\DeclareSymbolFont{mathb}{U}{mathb}{m}{n}
\DeclareMathSymbol{\mdownuparrows}{3}{mathb}{"D7}

\usepackage{makecell}

\usepackage{float}

\definecolor{dgreen}{rgb}{0,0.666,0}
\definecolor{dorange}{rgb}{0.666,.333,0}

\newcommand{\ie}{{\it i.e.},\ }

\newcommand{\Ad}{\operatorname{Ad}}
\newcommand{\ad}{\operatorname{ad}}

\makeatletter
\pgfmathdeclarefunction{erf}{1}{%
  \begingroup
    \pgfmathparse{#1 > 0 ? 1 : -1}%
    \edef\sign{\pgfmathresult}%
    \pgfmathparse{abs(#1)}%
    \edef\x{\pgfmathresult}%
    \pgfmathparse{1/(1+0.3275911*\x)}%
    \edef\t{\pgfmathresult}%
    \pgfmathparse{%
      1 - (((((1.061405429*\t -1.453152027)*\t) + 1.421413741)*\t 
      -0.284496736)*\t + 0.254829592)*\t*exp(-(\x*\x))}%
    \edef\y{\pgfmathresult}%
    \pgfmathparse{(\sign)*\y}%
    \pgfmath@smuggleone\pgfmathresult%
  \endgroup
}
\makeatother

\newcommand{\pd}{{\phantom{\dagger}}}
\renewcommand{\vec}[1]{\mathbf{#1}}

\begin{document}
\title{Majorana braiding simulations with projective measurements}

\author{Philipp Frey}
\affiliation{School of Physics, The University of Melbourne, Parkville, VIC 3010, Australia}

\author{Themba Hodge}
\affiliation{School of Physics, The University of Melbourne, Parkville, VIC 3010, Australia}

\author{Eric Mascot}
\affiliation{School of Physics, The University of Melbourne, Parkville, VIC 3010, Australia}

\author{Stephan Rachel}
\affiliation{School of Physics, The University of Melbourne, Parkville, VIC 3010, Australia}

\begin{abstract}
We summarize the key ingredients required for universal topological quantum computation using Majorana zero modes in networks of topological superconductor nanowires. Particular emphasis is placed on the use of both sparse and dense logical qubit encodings, and on the transitions between them via projective parity measurements. 
Combined with hybridization, these operations extend the computational capabilities beyond braiding alone and enable universal gate sets. 
In addition to outlining the theoretical foundations—including 
the algebra of Majorana operators, along with the stabilizer formalism—we introduce an efficient numerical method for simulating the time-dependent dynamics of such systems. 
This method, based on the time-dependent Pfaffian formalism, allows for the classical simulation of realistic device architectures that incorporate braiding, projective measurements, and disorder. 
The result is a semi-pedagogical overview and computational toolbox designed to support further exploration of topological quantum computing platforms.
\end{abstract}

\maketitle


\clearpage

\section{Introduction}




Topological quantum computing (TQC) provides a pathway to intrinsically fault-tolerant quantum computation by exploiting non-local encodings of quantum information in exotic quasiparticles such as Majorana zero modes (MZMs)~\cite{Nayak2008, Alicea2012, Aasen2016,ElliottFranz2015,Beenakker2013}. 
In particular, one-dimensional topological superconductors, as realized in the Kitaev wire model~\cite{Kitaev2001}, host spatially separated MZMs that can be braided to implement unitary gates protected from local decoherence~\cite{Ivanov2001, alicea2011}. 

Despite the elegance of topological protection offered by braiding, realizing a universal set of quantum gates within MZM-based architectures requires operations beyond braids. 
A central challenge lies in how logical qubits are encoded in collections of Majorana modes. 
Two primary encoding schemes have been proposed: \textit{sparse} and \textit{dense}~\cite{Karzig2017,Plugge2017}. 
Each presents distinct advantages and limitations.

In the \textit{sparse} encoding, each logical qubit is encoded with an ancillary pair of Majoranas to fix its total fermion parity~\cite{Karzig2017}. 
This setup permits the implementation of all single-qubit Clifford gates via braiding alone~\cite{Bravyi2006}. 
However, because braids preserve the parity of individual Majorana subsets, sparse encodings inherently forbid entangling operations between logical qubits. Any braid attempting to couple two sparse-encoded qubits violates their local parity constraints, thereby exiting the logical subspace. As a result, the sparse encoding cannot support multi-qubit logic gates solely through braiding~\cite{Karzig2017}.

Conversely, \textit{dense} encoding eliminates these local constraints by encoding multiple logical qubits into a minimal number of Majoranas plus a global parity-preserving ancilla~\cite{Karzig2017,Plugge2017}. This approach allows for entangling gates between qubits via braiding operations~\cite{Bravyi2006}. However, the absence of local parity constraints introduces a new obstacle: while two-qubit entangling gates can be realized, some single-qubit Clifford operations become inaccessible by braiding alone on all but a subset of qubits~\cite{Mascot2023}. Thus, dense encoding sacrifices complete local control for global entanglement.

To address these complementary limitations, this work focuses on a framework in which projective parity measurements enable dynamic transitions between sparse and dense encodings~\cite{Karzig2017,Plugge2017}. By temporarily relaxing local parity constraints through controlled measurements of joint Majorana parities (e.g., four-Majorana projectors), qubits can be coherently mapped between sparse and dense representations. This hybrid scheme permits single-qubit Clifford gates to be applied in sparse encoding, and then qubits can be mapped to dense encoding to perform entangling operations, such as CNOT or controlled-$Z$, before returning to sparse form. Crucially, the overall process preserves computational coherence, enabling the construction of universal gate sets~\cite{Karzig2017,Mascot2023}.

Additionally, we consider the role of \textit{Majorana hybridization}---the controlled energy splitting between two nearby Majorana modes~\cite{Hodge2025}. 
While braiding yields discrete gate operations (typically Clifford gates), hybridization allows the implementation of continuous single-qubit phase rotations, including T-gates, by dynamically controlling the overlap between Majorana wavefunctions~\cite{Bonderson2010,lahtinen-17spp021}. Hybridization complements the framework by enabling access to non-Clifford gates, which is essential for achieving universality with this approach~\cite{Bravyi2006,Mascot2023}.



This work presents an overview of a particular approach to universal topological quantum computation using MZMs, with an emphasis on classical simulation techniques. The ability to simulate systems that theoretically harbor MZMs across a wide range of parameters and including effects such as static and dynamic noise provides a crucial tool for the exploration of viable TQC platforms. It allows the user to set realistic expectations based on given system parameters, as well as determine practically workable parameter ranges that need to be realized in experiment.
In an accompanying work\,\cite{Hodge2025proj} we successfully demonstrate the simulation of up to 10 qubits, encoded in the joint state of 40 MZMs, performing universal quantum computation on a superconducting nanowire architecture.

In Chapter\,\ref{sec:Majoranas} we review the algebra of Majorana operators and their braiding statistics, and discuss the stabilizer formalism formulated in terms of these operators, as it is originally presented in\,\cite{Bravyi2006}. The latter is used to show why single encoding alone is insufficient for universality if we only allow for quadratic terms in the Hamiltonian. In Chapter\,\ref{sec:Sparse&Dense}, we review the sparse and dense encodings of logical qubits and describe how to implement transitions between them via joint-parity measurements, an essential ingredient for achieving universality\,\cite{Karzig2017}. Finally, Chapter\,\ref{sec:Simulation} develops a method for simulating the time-dependent dynamics of topological superconducting systems hosting MZMs using the Pfaffian formalism\,\cite{Wimmer2012, Mascot2023}. This includes the implementation of time-dependent projective measurements within the simulation framework.

By combining theoretical foundations with tractable simulation methods, this work aims to provide a self-contained resource for understanding and modeling universal topological quantum computation in systems based on Majorana zero modes.

\section{Majorana operators and braiding}\label{sec:Majoranas}

\subsection{Majorana operators}

The subspace $\mathcal{H}_0$ of zero-energy single-particle states can be characterized by a set of Hermitian Majorana operators, $\{\gamma_i\}$, which form a representation of the Clifford algebra,
\begin{equation}\label{eq:Clifford}
    \{ \gamma_i, \gamma_j \} = 2 \delta_{ij} \mathds{1}  \; ,
\end{equation}
where $\{ \cdot, \cdot \}$ denotes the anti-commutator \cite{Nayak2008,DasSarma2015,Field2018,BravyiKitaev2002,Vijay2015,Litinski2018}. Hence, $\gamma_i$ is also unitary, and $\gamma_i \gamma_j$ is skew-Hermitian for $i \neq j$. Any pair of Majorana operators $\gamma_i,\gamma_j$ with $i \neq j$ may be combined into a conventional Dirac fermion operator \mbox{$d_{(ij)}=(\gamma_i + i \gamma_j)/2$} that has the usual properties

\begin{equation}
\begin{aligned}
    \left(d_{(ij)}^{\vphantom{\dagger}}\right)^2 = \left(d_{(ij)}^\dagger\right)^2 &= 0 \\
    \left\{d_{(ij)}^\dagger, d_{(ij)}\right\} &= \mathds{1}  \\
    \left\{d_{(ij)}^\dagger, d_{(ij)}^\dagger\right\} = \left\{d_{(ij)}^{\vphantom{\dagger}}, d_{(ij)}\right\} &= 0  \;.
\end{aligned}
\end{equation}
The number operator $n_{ij}$ is given by

\begin{equation}
   n_{ij} = d_{(ij)}^\dagger d_{(ij)} = \frac{\mathds{1} + i \gamma_i \gamma_j}{2} \; ,
\end{equation}
where we made use of \eqref{eq:Clifford}. The operator $-i \gamma_i \gamma_j$ has the following properties:

\begin{align}
    \left(-i \gamma_i \gamma_j\right)^\dagger &= -i \gamma_i \gamma_j   \\
    \left(-i \gamma_i \gamma_j\right)^2 &= \mathds{1}  \\
    \gamma_i^\dagger \left(-i \gamma_i \gamma_j\right) \gamma_i &= + \left(i \gamma_i \gamma_j\right) \;.
\end{align}
The first property ensures that $i \gamma_i \gamma_j$ is diagonalizable and that all eigenvalues are real. The second property implies that they are equal to $\pm 1$. Finally, the last property follows from \eqref{eq:Clifford} and implies that there is an equal number of eigenvalues $+1$ and $-1$. The eigenvalues of the number operator $n_{ij}$ are therefore $0$ and $1$, as expected.
We refer to the operator $-i \gamma_i \gamma_j$ as the parity of the pair of Majoranas $\gamma_i, \gamma_j$ and states with eigenvalue $+1$\,($-1$) as even\,(odd) parity states.

A set of $2N$ Majorana operators can be partitioned into $N$ arbitrary pairs. The creation and annihilation operators $d_{(ij)}^\dagger, d_{(ij)}$ map between the two parities of the pair $\gamma_i, \gamma_j$ and commute with all other parity operators. Assuming that there exists a unique state $\ket{0}$ such that $ d_{(i_k j_k)}\ket{0}=0$ for all pairs $(i_k j_k)$ (no index repeated), then the subspace of degenerate ground states is spanned by the Fock states

\begin{equation}\label{eq:basis}
     \ket{n_1,\dots,n_N} = \left(d_{(i_1 j_1)}^\dagger\right)^{n_1} \cdots\left(d_{(i_N j_N)}^\dagger\right)^{n_N} \ket{0} \;,
\end{equation}
with $n_k = 0,1$ the eigenvalue of $n_{i_k j_k}$. The dimension of this subspace is therefore $\mathrm{dim}(\mathcal{H}_0) = 2^N$. Naively, this suggests that we have a register of $N$ qubits.

\subsection{Braiding}
In order to see how the information stored within the $N$ qubit system may be manipulated\,\cite{Ivanov2001,Kitaev2001}, we consider the time evolution of the system under (time-dependent) Hamiltonians that are quadratic in the Dirac fermion operators and, hence, also quadratic in the Majorana operators. The details of this time evolution depend on the system, but the resulting unitary may be implemented via an operator of the form $V(\theta)=\exp(\theta \gamma_i \gamma_j)$ with $i \neq j$ ($\gamma_i \gamma_j$ is skew-Hermitian). The effect on the information in the subspace spanned by $\{\gamma_i\}$ can be analyzed in the Heisenberg picture by considering the effect on Majorana operators, $\gamma_k^\prime = V^\dagger \gamma_k V$. We define the adjoint representation $\Ad$ of the Lie group U(N) on its Lie algebra of skew-Hermitian matrices $u(\mathrm{N})$, as well as the adjoint representation $\ad$ of the Lie algebra via

\begin{align}
    \Ad_V (x) &:= V x V^\dagger   \quad &&\textnormal{for} \quad V \in \mathrm{U(N)} ,\; x \in u(\mathrm{N}) \\
    \ad_x (y) &:= [x,y]           \quad &&\textnormal{for} \quad x,y  \in u(\mathrm{N}) \;,
\end{align}
where $[\cdot,\cdot]$ denotes the commutator. The Campbell identity relates the two representation:

\begin{equation}
    \Ad_{e^x}(y) = e^{\ad_x}(y) = y + [x,y] + \frac{1}{2!} [x,[x,y]] + \dots 
\end{equation}
We find that

\begin{equation}
\begin{aligned}
    \gamma_k^\prime &= \Ad_{V(-\theta)}(\gamma_k) =  e^{\ad_{V(-\theta)}}(\gamma_k)        \\
    &= \gamma_k - \theta [\gamma_i \gamma_j, \gamma_k] + \frac{1}{2} \theta^2 [\gamma_i \gamma_j,[\gamma_i \gamma_j, \gamma_k]] + \dots 
\end{aligned}
\end{equation}
The case where $i \neq k \neq j$ is trivial since any product of an even number of distinct Majorana operators commutes with any $\gamma_k$ that is not one of its factors and we find $\gamma_k^\prime = \gamma_k$. On the other hand, a product of an even number of distinct Majorana operators anti-commutes with any of its factors and so for $k=i,j$ we find 

\begin{equation}
     [\gamma_i \gamma_j, \gamma_k] = 
     \begin{cases}
         \hphantom{-} 2 \gamma_i, & \text{if } k=j \\
         -2 \gamma_j, & \text{if } k=i
     \end{cases} \;.
\end{equation}
The nested commutators are therefore cyclic and can be simplified to 

\begin{equation}\label{eq:general angle}
\begin{aligned}
     \gamma_i^\prime &= \left[\sum_n \frac{(2\theta)^{2n}}{(2n)!} (-1)^n \right] \gamma_i -  \left[\sum_n \frac{(2\theta)^{2n+1}}{(2n+1)!} (-1)^n \right] \gamma_j  \\
    &= \cos(2\theta) \gamma_i - \sin (2\theta) \gamma_j    \\
    \\
    \gamma_j^\prime &= \sin (2\theta) \gamma_i + \cos(2\theta) \gamma_j \;,
\end{aligned}
\end{equation}
where we have omitted the series representation for the second equality. This is simply the Givens rotation matrix $G(i,j,2\theta)$.
The special case $\theta = \pi/4$ clearly exchanges the two Majoranas up to a sign and represents a braiding operation

\begin{equation}\label{eq:braid}
\begin{aligned}
    B_{ij} :&= \exp \left[ \frac{\pi}{4} \gamma_i \gamma_j \right] = \frac{1}{\sqrt{2}}(\mathbb{1}+\gamma_i \gamma_j) \\
    B^\dagger_{ij} \gamma_i B_{ij} &= -\gamma_j    \\
    B^\dagger_{ij} \gamma_j B_{ij} &= \hphantom{-}\gamma_i \\
    B^\dagger_{ij} \gamma_k B_{ij} &= \hphantom{-}\gamma_k  \quad ,i \neq k \neq j \;.
\end{aligned}
\end{equation}
For reference, we note the following identities defined for the nearest-neighbor braids $\bar{B}_{i}:=B_{i,i+1}$, which form a representation of the Artin braid group:
\begin{equation}
\begin{aligned}
    \bar{B}_{i} \bar{B}_{j} &= \bar{B}_{j} \bar{B}_{i} &&,|i-j|>1   \\
    \bar{B}_{i} \bar{B}_{j} \bar{B}_{i} &= \bar{B}_{j} \bar{B}_{i} \bar{B}_{j}  &&,|i-j|=1 \;.   \\
\end{aligned}
\end{equation}
In order to see the effect of braids on the $N$-qubit state we make use of the stabilizer formalism.


\subsection{Stabilizer formalism}

The central idea of the stabilizer formalism \cite{Gottesman1997,Mudassar2024, AaronsonGottesman2004,Gottesman1998,BravyiKitaev2005,Bravyi2006,Bravyi2010} is that an element of the Hilbert space may be determined by a specifying a set of operators that act trivially on said element. Consider a group $G$ that acts on some set $X$ via a group action $G \times X \to X, (g,x) \mapsto gx$. In our case $G \subset \mathrm{U(n)}$ will be unitary operators that are also Hermitian and $X=\mathcal{H}_0$ the Hilbert space spanned by \eqref{eq:basis}. A stabilizer of $x \in X$ is a group element $g \in G$ that leaves $x$ invariant, \ie, $gx=x$. The set $G_x=\{g \in G | gx=x \}$ is clearly a subgroup of $G$ and is called the stabilizer subgroup with respect to $x$. In the context of topological quantum information we will consider $G$ to be the group $G=\{\alpha \gamma_{i_1} \gamma_{i_2} \cdots \gamma_{i_m} | \alpha \in \{\pm 1,\pm i\}, m \in \mathbb{N} \}$, \ie, any finite product of Majorana operators with a pre-factor of either $\pm 1$ or $\pm i$. This is clearly a finite group due to \eqref{eq:Clifford}. In particular, $G$ and any of its subgroups such as $G_x$ are finitely generated, meaning there exist a finite number of group elements such that $G_x = \langle g_1, \dots,g_n \rangle$. Here, $\langle \cdot \rangle$ denotes the set of all finite products of the elements $g_1, \dots, g_n$. The concept of stabilizers is particularly useful in the case where specifying $G_x$ uniquely determines $x$. To see how this works we return to the case of the Fock space defined by $2N$ Majorana operators \eqref{eq:basis}. The Fock states are uniquely characterized (up to an irrelevant phase) by their eigenvalue with respect to all parity operators (stabilizers) $i \gamma_{i_k} \gamma_{j_k}$. Therefore, $\ket{n_1,\dots,n_N}$ is stabilized by the group $\langle (-1)^{n_1} i \gamma_{i_1} \gamma_{j_1} \;, \dots, \; (-1)^{n_N} i \gamma_{i_{N}} \gamma_{j_N} \rangle$ and the relation is one-to-one.

A unitary operation $V$ on the state can be expressed in terms of its action on the set of stabilizers \mbox{$g_i \mapsto g^\prime_i = V^\dagger g_i V$}, where $g^\prime_i$ is a stabilizer of the state $V \ket{\psi}$ iff $g_i$ is a stabilizer of $\ket{\psi}$. For simplicity we choose the initial state $\ket{0}$ with respect to a particular pairing of Majoranas and consider the effect of braid operations. Let the initial complete set of stabilizers be given by

\begin{equation}
    (-i \gamma_1 \gamma_2), (-i \gamma_3 \gamma_4), \dots, (-i \gamma_{2N-1} \gamma_{2N}) \;.
\end{equation}
The action of $B_{ij}$ only changes stabilizers that involve $\gamma_i$ or $\gamma_j$. If $i,j$ belong to the same stabilizer then it follows from \eqref{eq:braid} that all stabilizers are left invariant, \ie, 

\begin{equation}\label{eq:stabilizer braid 1}
    B^\dagger_{ij} (\pm i \gamma_i \gamma_j) B_{ij} = (\mp i \gamma_j \gamma_i) =  (\pm i \gamma_i \gamma_j)  \;.
\end{equation}
We note that knowledge of the stabilizers does not determine the phase of the state. Indeed, $B_{ij}$ will change the phase by $e^{\pm i \pi/4}$ and therefore acts like a logical $\sqrt{Z}$ gate on the qubit defined by $(ij)$.
For the case where $i,j$ belong to different stabilizers we consider w.l.o.g. the braid $B_{23}$ which maps 

\begin{equation}\label{eq:stabilizer braid 2}
\begin{aligned}
    -i \gamma_1 \gamma_2 \quad \overset{B_{23}}{\mapsto} \quad \hphantom{-}i \gamma_1 \gamma_3  \quad \overset{B_{23}}{\mapsto} \quad i \gamma_1 \gamma_2 \\
    -i \gamma_3 \gamma_4 \quad \overset{B_{23}}{\mapsto} \quad -i \gamma_2 \gamma_4  \quad \overset{B_{23}}{\mapsto} \quad i \gamma_3 \gamma_4
\end{aligned}    
\end{equation}
Performing two braids returns the same set of stabilizers except with opposite signs, corresponding to a mapping $\ket{00} \mapsto e^{i \phi} \ket{11}$. A single braid, on the other hand, gives a different pairing of Majorana operators. Using $\gamma_{2n-1}=(d_n^\dagger+d_n)$, $\gamma_{2n} = i(d_n^\dagger-d_n)$ we can express $i \gamma_1 \gamma_3$ and $-i \gamma_2 \gamma_4$ the original basis and find the common +1 eigenstate, which is $(\ket{00}+i\ket{11})/\sqrt{2}$. A single braid has entangled the two logical qubits. In general, a sequence of braids will map the initial set of stabilizers by permuting the Majorana indices, changing signs of parities and possibly multiplying the state by a phase factor:

\begin{equation}\label{eq:stabilizer perm}
    \begin{aligned}
         &(-i \gamma_1 \gamma_2) \;, \dots, \; (-i \gamma_{2N-1} \gamma_{2N})  \\
         \mapsto \quad & (\pm i \gamma_{\sigma(1)} \gamma_{\sigma(2)}), \dots, (\pm i \gamma_{\sigma(2N-1)} \gamma_{\sigma(2N)})
    \end{aligned}
\end{equation}
Similar to the parity $i \gamma_i \gamma_j$ of a single pair of Majoranas, we can also define the total parity of any even number of Majoranas as the product of the parities of individual pairs, $\prod_{k=1}^n (i \gamma_{i_k} \gamma_{j_k}) = \alpha \gamma_{i_1}\cdots\gamma_{j_n}$, where $\alpha \in \{ \pm 1, \pm i \}$ and no index is repeated. This operator clearly has eigenvalues $+1\,(-1)$ corresponding to an even (odd) total number of fermions. However, the individual parities of all pairs of Majoranas need not be well defined. 
A braid that involves two Majoranas that are factors of the same parity operator will subsequently commute with said operator.
This follows from \eqref{eq:braid} and the fact that the permutation of $\gamma_i$ and $\gamma_j$ can be undone by $m + (m-1)$ anti-commutations, which corresponds to an odd number of sign flips that cancels the sign in \eqref{eq:braid}. The conservation of total parity implies that, starting from the vacuum state $\ket{0}$, only half of all states $\ket{n_1,\dots,n_N}$ can be reached by braiding. The number of logical qubits realized by a system of $2N$ Majoranas is therefore at most $N-1$.

In the following section, we will introduce two different approaches to encoding $N$ logical qubits in terms of $2N^\prime$ Majoranas and discuss their advantages and limitations.

\section{Sparse and dense encoding}\label{sec:Sparse&Dense}

As we have shown above, $N$ logical qubits cannot be encoded in $2N$ Majorana operators since the conservation of total fermion parity reduces the accessible Hilbert space dimension by a factor of 2. This can be circumvented by adding ancilla qubits in the form of additional pairs of Majoranas to the system. However, they do not represent logical qubits but serve to absorb excess parity during braiding operations. 

\subsection{Sparse encoding}
\begin{figure}[b!]
    \centering
    \includegraphics[scale=0.82]{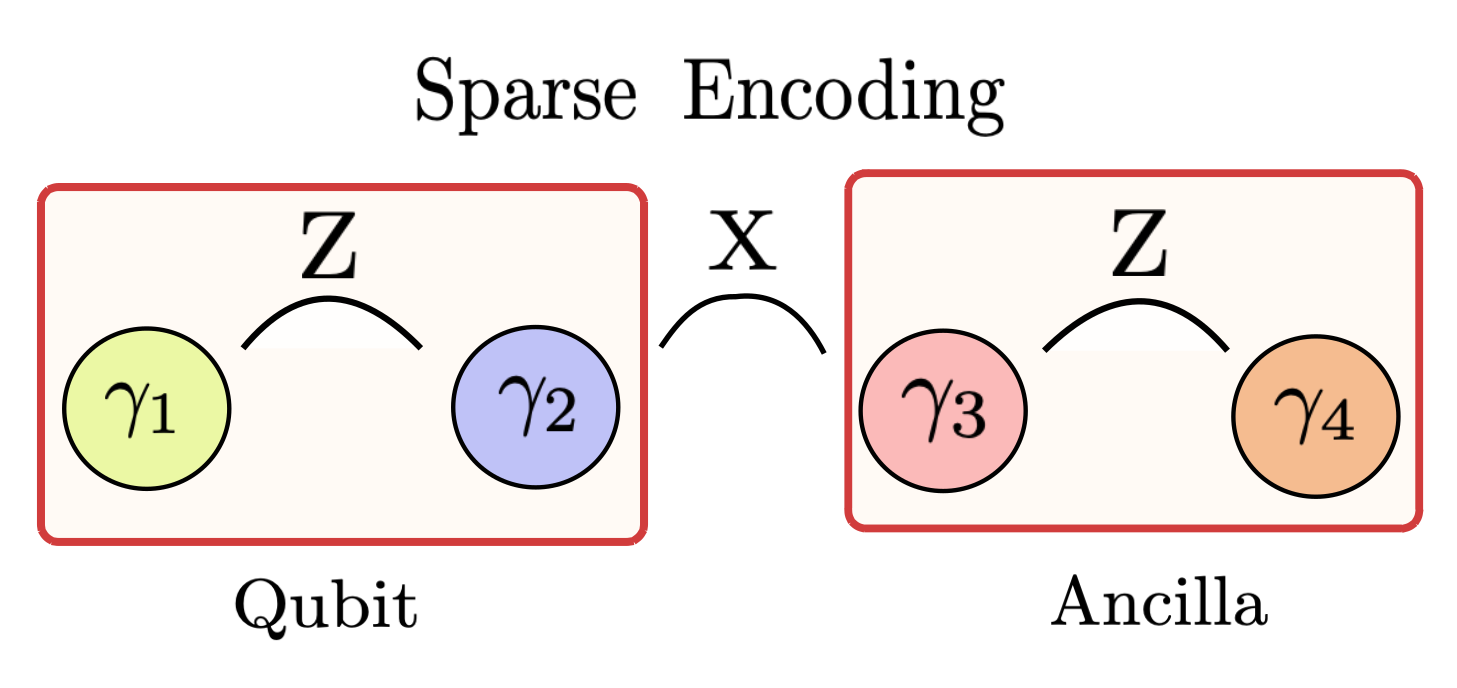}
    \caption{Schematic of the sparse encoding for a single qubit, with the logical qubit labelled along with the parity conserving ancilla. Hybridization between pairs of Majoranas results in continuous $X$- or $Z$-rotation around the Bloch sphere as indicated.}
    \label{fig:Fig1}
\end{figure}
One possible approach is to provide every logical qubit with its own ancilla qubit (see Fig.\,\ref{fig:Fig1}), which means that $4N$ Majoranas encode $N$ logical qubits \cite{Bravyi2006,DasSarma2015,Bonderson2008,Karzig2017}. In order to project the physical Hilbert space of dimension $\operatorname{dim}(\mathcal{H}_0) = 2^{2N}$ onto a logical \mbox{$2^N$-dimensional} subspace, constraints have to be imposed. Fixing the combined parity of each qubit and its ancilla to be even provides $N$ independent constraints, each of which removes half of the basis states,

\begin{equation}
    - \gamma_{4i-3} \gamma_{4i-2} \gamma_{4i-1} \gamma_{4i} \ket{\psi} = \ket{\psi}  \quad , i = 1,\dots,N .
\end{equation}
For a single qubit defined by 4 Majoranas, the logical basis states are given by $\ket{0}_\mathrm{log}=\ket{00}$ and $\ket{1}_\mathrm{log}=\ket{11}$ in the $(-i \gamma_1 \gamma_2)$, $(-i \gamma_3 \gamma_4)$ basis. If we consider braiding of Majoranas that make up a single qubit, we can refer to \eqref{eq:stabilizer braid 1}, \eqref{eq:stabilizer braid 2}, and the comments thereafter to see that braiding either $\gamma_1, \gamma_2$ or, equivalently, $\gamma_3, \gamma_4$ implements a $\sqrt{Z}$-gate. A braid between operators $\gamma_2, \gamma_3$ corresponds to a logical $\sqrt{X}$-gate, since we have seen that $B_{23}\ket{0}_\mathrm{log} = (\ket{0}_\mathrm{log}+i \ket{1}_\mathrm{log})/\sqrt{2}$. Combined with the relations \mbox{$\sqrt{Y} \propto \sqrt{X}^{-1}\sqrt{Z}\sqrt{X}$} and \mbox{$H \propto \sqrt{X}\sqrt{Z}\sqrt{X}$} we find that we can implement any single-qubit Clifford gate via braiding. To achieve universal quantum computation this gate set would have to be extended to include, say, arbitrary phase gates $R_z(\phi) = \exp (-i \phi Z/2)$ with arbitrary angle $\phi$. This might be achievable if one makes use of hybridization, as we discuss below.

The problem with the sparse encoding is the inability to entangle logical qubits via braid operations. To see this, consider that braids between arbitrary qubits map the stabilizers according to \eqref{eq:stabilizer perm}. A stabilizer $(\pm i \gamma_{\sigma(2i-1)} \gamma_{\sigma(2i)})$ commutes with the constrained parity $p_j = - \gamma_{4j-3} \gamma_{4j-2} \gamma_{4j-1} \gamma_{4j}$ iff either both $\gamma_{\sigma(2i-1)}, \gamma_{\sigma(2i)}$ are factors of $p_j$, or neither of them. Otherwise, the two parities anti-commute and cannot simultaneously be equal to $+1$ on the state defined by the stabilizer group, and therefore the state is not in the logical subspace. Hence, any braid that maps back into the logical subspace must pair the four Majoranas that define a given qubit into two pairs that make up stabilizers. In other words, no stabilizers can involve a product of Majorana operators that belong to different logical qubits. Each logical qubit in the sparse encoding is comprised of two qubits and we have just proven that the corresponding four Majorana operators have to appear in two generators of the stabilizer group. Their joint eigenstate with eigenvalue $+1$ is found as a superposition of the four Fock states of the two qubits, which is a single-qubit state with respect to the logical qubit. This applies to all logical qubits, and hence the joint eigenstate with eigenvalue $+1$ is simply a product state over the individual logical qubits.
We therefore find that no entanglement between logical qubits can result from braiding operations.

The same applies to any encoding that introduces parity constraints on subsets of Majorana operators, where entanglement across separately constrained subsets is ruled out.  

\subsection{Dense encoding}

\begin{figure}[b!]
    \centering
    \includegraphics[scale=0.85]{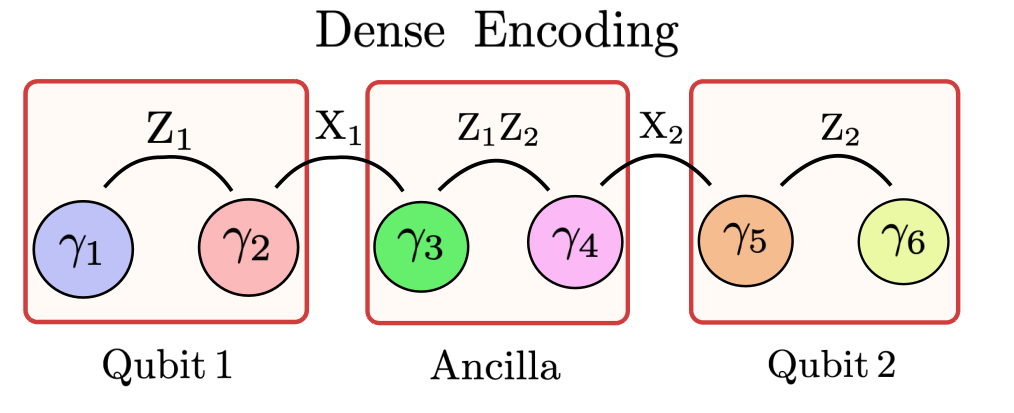}
    \caption{Schematic of the dense encoding for two qubits, with both logical qubits labelled along with the parity conserving ancilla. Hybridization between pairs of Majoranas causes rotations in the logical code space as indicated.}
    \label{fig:Fig2}
\end{figure}

We conclude based on the analysis in the previous section that the only possible encoding that could allow for arbitrary entanglement between logical qubits is one that has no additional constraints apart from the necessary conservation of total parity for the whole physical system.
Naturally, we are led to what is called the dense encoding of $N$ logical qubits in $2(N+1)$ physical Majoranas (see Fig.\,\ref{fig:Fig2}), whose accessible Hilbert space has the right dimension of $2^{(N+1)}/2$ \cite{Bravyi2006,Karzig2017,Plugge2017,DasSarma2015}. One pair of Majoranas forms an ancilla qubit that can absorb parity and potentially allows for arbitrary single-qubit rotations. In order to see what gates can and cannot be implemented via braiding, we make use of a representation of the Majoranas in terms of Pauli operators $P_i \in \{X_i,Y_i,Z_i \}$ on $N+1$ qubits. Here, $P_i$ is short for $I \otimes \cdots \otimes I \otimes P \otimes I \otimes \cdots \otimes I $, with $P$ a Pauli operator at the $i$'th position. A Jordan-Wigner transformation makes operators on different sites anti-commuting and allows us to define

\begin{equation}
\begin{aligned}
    \gamma_{2i-1} &:= \left( \prod_{j<i} Z_j \right) X_i \\
    \gamma_{2i}   &:= \left( \prod_{j<i} Z_j \right) Y_i        \;.
\end{aligned}
\end{equation}
One can readily verify that \eqref{eq:Clifford} is satisfied and furthermore that $-i \gamma_{2i-1} \gamma_{2i}=Z_i$, meaning the Pauli operators coincide with the logical $(X,Y,Z)$ gates on qubit $i$. The initial set of stabilizers for the $\ket{0}$ state are therefore 

\begin{equation}
    Z_1, Z_2, \dots, Z_N, Z_{N+1}   \;.
\end{equation}
The total parity is simply the product $Z_1 Z_2 \cdots Z_{N+1}$ and commutes with any sequence of braids, \ie, is conserved. The braids of \eqref{eq:stabilizer braid 2} expressed in Pauli operators are 

\begin{equation}\label{eq:stabilizer braid spin}
\begin{aligned}
    &\pm Z_1  &&\overset{B_{23}}{\mapsto} \quad \pm Y_1 X_2  \quad \overset{B_{23}}{\mapsto} \quad  \mp Z_1 \\
    &\pm Z_2  &&\overset{B_{23}}{\mapsto} \quad \pm X_1 Y_2  \quad \overset{B_{23}}{\mapsto} \quad \pm Z_2 \\
    &\pm Z_i  &&\overset{B_{23}}{\mapsto} \quad \pm Z_i  \quad , i=3,\dots,N+1      \;.
\end{aligned}    
\end{equation}
If we take the first  qubit to be the ancilla qubit and qubits  2 to $N+1$ to be the logical qubits 1 to $N$, then the first braid $B_{23}$ in \eqref{eq:stabilizer braid spin} maps 

\begin{equation}
    \begin{aligned}
        &\ket{0 \, 0 \, n_2 \cdots n_N} \mapsto \frac{1}{\sqrt{2}} \left( \ket{0 \, 0 \, n_2 \cdots n_N} + i \ket{1 \, 1 \, n_2 \cdots n_N} \right)  \\
        &\ket{1 \, 1 \, n_2 \cdots n_N} \mapsto \frac{1}{\sqrt{2}} \left( \ket{0 \, 0 \, n_2 \cdots n_N} - i \ket{1 \, 1 \, n_2 \cdots n_N} \right)  \\
        &\ket{0 \, 1 \, n_2 \cdots n_N} \mapsto \frac{1}{\sqrt{2}} \left( \ket{1 \, 0 \, n_2 \cdots n_N} - i \ket{0 \, 1 \, n_2 \cdots n_N} \right)  \\
        &\ket{1 \, 0 \, n_2 \cdots n_N} \mapsto \frac{1}{\sqrt{2}} \left( \ket{1 \, 0 \, n_2 \cdots n_N} + i \ket{0 \, 1 \, n_2 \cdots n_N} \right)   \;,
    \end{aligned}
\end{equation}
where the $n_2 \cdots n_N$ are such that the overall parity is even but otherwise arbitrary and unaffected by the operation. This can be seen by explicitly verifying that these are the eigenstates (up to a phase) of the corresponding stabilizers with eigenvalue one. While this leaves the possibility of additional relative phases, explicit calculation using $B_{23}=\exp[i \pi/4 X_1 X_2]$ shows that there are none, and we choose the stabilizer representation here for the purpose of contrasting this result with other braids below. By direct comparison with the effect of a logical $\sqrt{X}$ rotation on qubit 1, along with the necessary change of the state of the ancillary to preserve total parity, it is clear that the braid $B_{23}$ affects exactly that operation. In similar fashion, it can be seen that $B_{34}$ corresponds to a $\sqrt{Z}$ rotation. We therefore have all Clifford gates on the first qubit. Similarly, we can consider braiding one of the ancilla Majoranas with one of the Majoranas of qubit $m$ in order to try and implement a single-qubit rotation on this qubit. All other stabilizers should be the same before and after the braiding to conserve the state of the other logical qubits. This restricts the resulting set of generators $\{S_i\}$ of the stabilizer group to be of the form

\begin{equation}\label{eq:stabilizer braid m'th spin}
\begin{aligned}
    &\pm Z_1  &&\mapsto \ \pm Y_1 Z_2 \cdots Z_{m-1} X_{m}  \ \text{or} \  \pm X_1 Z_2 \cdots Z_{m-1} X_{N+1}\\
    &\pm Z_m  &&\mapsto \ \pm X_1  Z_2 \cdots Z_{m-1} Y_m  \ \text{or} \  \pm Y_1 Z_2 \cdots Z_{m-1} Y_{N+1} \\
    &\pm Z_i      &&\mapsto \ \pm Z_i  \quad , i \neq m       \;.
\end{aligned}    
\end{equation}
Note that the signs are not independent since \mbox{$\prod_i S_i = \prod_i Z_i$} due to the parity preserving property of braids. For simplicity we will only consider one of the possible braids hereafter, since all arguments hold for either case.
For example, if we braid $\gamma_2$ and $\gamma_{2N+1}$ the stabilizers transform according to

\begin{equation}\label{eq:stabilizer braid last spin}
\begin{aligned}
    &\pm Z_1  &&\overset{B_{1 \, 2N+1}}{\mapsto} \quad S_1 = \pm Y_1 Z_2 \cdots Z_{N} X_{N+1}  \\
    &\pm Z_{N+1}  &&\overset{B_{1 \, 2N+1}}{\mapsto} \quad S_2 = \pm X_1  Z_2 \cdots Z_{N} Y_{N+1} \\
    &\pm Z_i      &&\overset{B_{1 \, 2N+1}}{\mapsto} \quad S_i = \pm Z_i  \quad , i=2,\dots,N        \;.
\end{aligned}    
\end{equation}
In general, any product of the form $\prod_{j \in J} S_j$, where \mbox{$J \subseteq \{1,\dots,N\}$}, is also a stabilizer and may replace one of its factors without losing the generating property of the set, since all stabilizers are their own inverse. The first two transformations in \eqref{eq:stabilizer braid last spin} suggest that the action on the last qubit depends on the state of all other logical qubits, as the factors $Z_i$ can only be undone by multiplication with the stabilizer $\pm Z_i$ and therefore may introduce a sign change depending on the parity of these qubits. However, in this case we can make use of the total parity $P=Z_1 \cdots Z_{N+1} = \prod_j S_j$ being a stabilizer. We find that $S_1 P = \prod_{j\neq 1} S_j$ may replace $S_2$ and $S_2 P = \prod_{j\neq 2} S_j$ may replace $S_1$:

\begin{equation}
\begin{aligned}
    & S_1 =\pm Y_1 Z_2 \cdots Z_{N} X_{N+1} \quad \mapsto \quad \pm Y_1 X_{N+1} \\
    & S_2 = \pm X_1 Z_2 \cdots Z_{N} Y_{N+1} \quad \mapsto \quad  \pm X_1 Y_{N+1} \;.
\end{aligned}    
\end{equation}
Importantly, the signs are unchanged in this special case.
Comparison with \eqref{eq:stabilizer braid spin} shows that this is indeed a $\sqrt{X}$-gate on the last qubit. However, for all logical qubits, except the first and last one, this does not work, since the only product of generators that removes the Jordan-Wigner string in between the two $X(Y)$ operators is given by $\prod_{1<j<m} \pm Z_j$. For $m \neq 2,N+1$, there is no constraint on the overall sign in this case. The stabilizers after the corresponding braids are of the form

\begin{equation}\label{eq:stabilizer braid bulk spin}
\begin{aligned}
    &\pm Y_1 Z_2 \cdots Z_{m-1} X_m       \quad \simeq \quad \pm p(2,\dots,m-1) \, Y_1  X_m      \\
    &\pm X_1  Z_2 \cdots Z_{m-1} Y_m      \quad \simeq \quad \pm  p(2,\dots,m-1) \, X_1  Y_m           \\
    &\pm Z_j  \qquad \qquad \quad ,\, j \neq m                 \;,
\end{aligned}    
\end{equation}
where $p(2,\dots,m-1) = \prod_{j=2}^{m-1} (-1)^{n_j} = \pm 1$ is the total parity of all qubits between the ancilla and its braiding partner. The first two stabilizers in \eqref{eq:stabilizer braid bulk spin} were multiplied by $\prod_{1<j<m} \pm Z_j$ to arrive at the final results. We see that the logical operation is not disentangled from the state of all other qubits and therefore is not equivalent to a pure single-qubit rotation.

We have seen that braids between Majorana operators from different qubits such as $B_{2j,2j+1} = \exp[i \pi/4 X_j X_{j+1}]$ act as joint rotations on these qubits. For the case of two different logical qubits, this is an entangling gate. Hence we conclude that the dense encoding provides no obstacle to entangling any two logical qubits via braiding, but does not allow for even the Clifford gates on all qubits. Adding Majorana hybridization on top of braids does not resolve this issue, as we will now discuss.

\subsubsection{Hybridization}

We now consider an extension of the braid group by making use of the hybridization between Majoranas\,\cite{Kitaev2001,Ivanov2001,Nayak2008,alicea2011,Albrecht2016,Plugge2017,Karzig2017}. While the zero-energy space $\mathcal{H}_0$ is exactly degenerate in the limit where all Majorana particles are infinitely distant from each other, this degeneracy is generally lifted when the separation is finite. In particular, when $\gamma_i$ and $\gamma_j$ are close to each other their even and odd parity states are split by energy $\epsilon_{ij}$. This means that a dynamical phase of $e^{\epsilon_{ij} \gamma_i \gamma_j t}$ is acquired over a time $t$ \cite{Bonderson2010,cheng2011,Hodge2025}. In a realistic scenario, $\epsilon_{ij}(t)$ will be time dependent and the phase angle is given by $\int \epsilon_{ij}(t') dt'$. We will now assume that hybridization between any pair of Majoranas can be controlled such that arbitrary angles may be realized. The operator $e^{\theta_{ij} \gamma_i \gamma_j}$ generalizes braiding operations, which correspond to the braid angle $\theta_{ij}=\pi/4$. Eq.~\eqref{eq:general angle} shows that the action on Majorana operators is a rotation in the $(i,j)$-plane by an angle $2\theta_{ij}$, with braids being special case of $90^{\circ}$ rotations. In fact, the set of all bilinears $\gamma_i \gamma_j$ with $i \neq j$ form a representation of the Lie algebra $\mathfrak{so}(2(N+1))$, \ie, generators of rotations in $2(N+1)$ dimensions. Any product of operators of the form $\prod_{i \neq j} e^{\theta_{ij} \gamma_i \gamma_j} =  e^{\chi}$ may be expressed as an exponential of a general Lie-algebra element $\chi = \sum_{i \neq j} \theta'_{ij} \gamma_i \gamma_j$. We can therefore characterize the group of transformations on the logical Hilbert space by studying the action of its Lie algebra. First, we note that $\mathcal{D}_{\text{maj}} := \operatorname{dim}(\mathfrak{so}(2(N+1))) = \binom{2(N+1)}{2}=(2N+1)(N+1)$, whereas $\mathcal{D}_{\text{uni}} :=\operatorname{dim}(\mathfrak{su}(2^N)) = 4^N-1$, the latter being the target if universality is to be achieved. 

For $N=1$, we have $\mathcal{D}_{maj}=6$ and $\mathcal{D}_{uni}=3$. This is due to the projection of the physical Hilbert space onto the even-parity subspace, where braiding the ancilla Majoranas, for example, is equivalent to braiding or hybridizing the qubit Majoranas. Both implement the same $Z$-rotation in the logical subspace. For $N=1$, braiding plus hybridization is universal. 

For $N=2$ we find $\mathcal{D}_{maj}=\mathcal{D}_{uni}=15$, meaning that we have universality as well. This may be understood based on our earlier finding that $\sqrt{X}$-gates can be implemented via braiding on qubits $1$ and $N-1$, which, together with the always available  $\sqrt{Z}$-gates gives us Clifford gates on both logical qubits. Furthermore, as we will see below, a CNOT-gate between the two can be expressed via braids in this case, which completed the Clifford group. Finally, hybridization adds a $T$-gate and therefore gives us universality.

For $N>2$ we have $\mathcal{D}_{maj} < \mathcal{D}_{uni}$, which rules out universality, and thus also the possibility that all Clifford gates may be realized when hybridization is added to the gate set. We can go further and rule out the possibility that $\sqrt{X}$-gates on all logical qubits could be realized. If this were the case, then combined with the $\sqrt{Z}$- and $T$-gates we would have a universal gate set for the single-qubit unitaries. A single braid of Majoranas between neighboring qubits is an entangling gate, as we have seen earlier. It is known that the combination of all single-qubit unitaries plus a single entangling gate acting on nearest neighbors forms a universal gate set, which leads to a contradiction.


\begin{figure}[b!]
    \centering
    \includegraphics[scale=1.0]{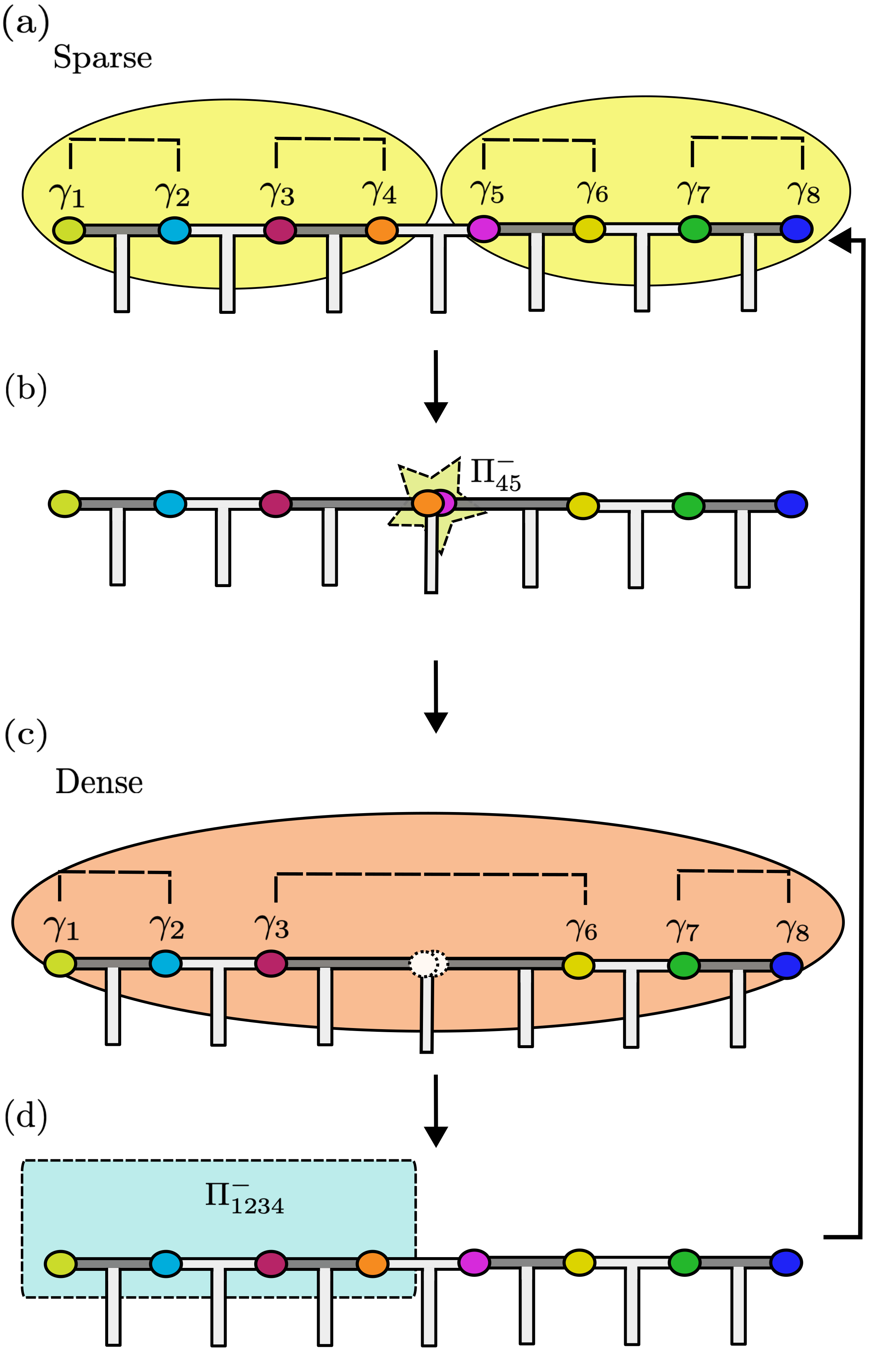}
    \caption{Schematic of the sparse to dense projection process.
    (a) Sparse encoding for two logical qubits each highlighted in yellow. 
    (b) Illustration of the $\Pi^-_{45}$ projection, with the measurement conducted via pairwise fusion, mapping the qubit to the dense encoding. 
    (c) Resulting dense encoding of the two logical qubits in terms of six MZMs highlighted in orange. Note that $\gamma_4$, $\gamma_5$ are absent from the encoding. 
    (d) Here we highlight the MZMs involved in the four-point $\Pi^-_{1234}$ projection, which maps the qubit back to the sparse encoding. } 
    \label{fig:Fig3}
\end{figure}

\subsection{Mapping between sparse and dense encoding}

We have seen that the sparse encoding allows for all single-qubit Clifford gates, with the possibility of extension to a complete gate set if hybridization is added. Logical qubits cannot be entangled, however. In contrast, in the dense encoding, qubits are readily entangled, but not all single-qubit Clifford gates are available. A possible solution is to use both encodings by mapping back and forth between them\,\cite{DasSarma2015,zhan_dissipationless_2024,zhan_universal_2022}. Single-qubit gates are implemented in the sparse encoding, while entangling gates are implemented in the dense encoding. To see how this is achieved, consider a set of 8 Majoranas that define two logical qubits in the sparse encoding (cf.\ Fig.\,\ref{fig:Fig3}). We define the following operators

\begin{equation}
    \begin{aligned}
        Q_{s,1} &:= - \gamma_1 \gamma_2 \gamma_3 \gamma_4    \\
        Q_{s,2} &:= - \gamma_5 \gamma_6 \gamma_7 \gamma_8    \\ 
        Q_{d\hphantom{,1}}     &:= i \gamma_1 \gamma_2 \gamma_3 \gamma_6 \gamma_7 \gamma_8  \\
        \Pi^\mp_{45}  &:= \frac{1}{2} \left( \mathbb{1} \mp i \gamma_4 \gamma_5 \right)         \\
        \Pi^\mp_{1234}  &:= \frac{1}{2} \left( \mathbb{1} \mp \gamma_1 \gamma_2 \gamma_3 \gamma_4 \right) \;. \label{eq:Projectors}
    \end{aligned}
\end{equation}
$Q_{s,1}, Q_{s,2}$ and $Q_{d}$ are simply the total parity of the respective Majorana operators from which they are constructed. A state $\ket{\psi}_s$ in the logical subspace of the sparse encoding is defined by the condition $Q_{s,1}\ket{\psi}_s = Q_{s,2}\ket{\psi}_s=\ket{\psi}_s$. On the other hand, a state $\ket{\psi}_d$ in the logical subspace of the dense encoding defined by Majoranas $1,2,3,6,7,8$ has the defining property \mbox{$Q_{d}\ket{\psi}_d=\ket{\psi}_d$}. The operators $\Pi^\mp_{45},\Pi^\mp_{1234}$ are pairwise orthogonal projection operators, meaning they are Hermitian and satisfy

\begin{equation}
    \begin{aligned}
        (\Pi^\mp_{45})^2 = \Pi^\mp_{45} \quad &, \quad (\Pi^\mp_{1234})^2 =\Pi^\mp_{1234}    \\
        \Pi^-_{45} + \Pi^+_{45} = \mathbf{1}  \quad &, \quad  \Pi^-_{1234} + \Pi^+_{1234} = \mathbf{1}  \\
        \Pi^-_{45} \Pi^+_{45} =0  \quad &, \quad   \Pi^-_{1234} \Pi^+_{1234} =0 
    \end{aligned}
\end{equation}
Starting with a logical state $\ket{\psi}_s$ in the sparse encoding we map to the dense encoding by measuring the parity of $\gamma_4, \gamma_5$. This may be done by fusing and measuring the resulting fermion. Assuming we measure no fermion, this process corresponds to $\ket{\psi}_s \mapsto \ket{\psi'} = \mathcal{N'} \, \Pi^-_{45} \ket{\psi}_s$, with $\mathcal{N'}$ a normalization constant to be determined. We can verify that $\ket{\psi'}$ is a logical state in the dense encoding by first rewriting 

\begin{equation}
    \begin{split}
        Q_d &= i \gamma_1 \gamma_2 \gamma_3 \gamma_4^2 \gamma_5^2 \gamma_6 \gamma_7 \gamma_8 \\
            &= -i \gamma_1 \gamma_2 \gamma_3 (\gamma_4 \gamma_5) \gamma_4 \gamma_5 \gamma_6 \gamma_7 \gamma_8  \\
            &= -i (\gamma_4 \gamma_5) \gamma_1 \cdots \gamma_8  \\
            &= (-i \gamma_4 \gamma_5) Q_{s,1} Q_{s,2} \;,
    \end{split}
\end{equation}
which implies that, applied to $\ket{\psi}_s$, we have the identity

\begin{equation}
    Q_d \ket{\psi}_s = -i \gamma_4 \gamma_5 \ket{\psi}_s    \;.
\end{equation}
Combined with the vanishing commutator, $[Q_d,\gamma_4 \gamma_5]=0$, it follows that

\begin{equation}\label{eq:sparse-to-dense}
    \begin{split}
        Q_d \ket{\psi'} &= \frac{1}{2} \left( \mathbb{1} - i \gamma_4 \gamma_5 \right) Q_d  \ket{\psi}_s \\
                        &= \frac{1}{2} \left(-i \gamma_4 \gamma_5 + (i \gamma_4 \gamma_5)^2 \right) \ket{\psi}_s  \\
                        &= \Pi^-_{45} \ket{\psi}_s = \ket{\psi'} /\mathcal{N'}     \;,
    \end{split}
\end{equation}

so $\ket{\psi'}$ is in the logical subspace of the dense encoding. Starting from a state $\ket{\psi}_d$, in the dense encoding, we initialize a pair of Majoranas $\gamma_4, \gamma_5$ in the even parity state, \ie, $Q_d \ket{\psi}_d = -i \gamma_4 \gamma_5 \ket{\psi}_d = \ket{\psi}_d$, and measure the combined parity of $1,2,3,4$. If the outcome is even, this corresponds to the mapping $\ket{\psi}_d \mapsto \ket{\psi''} = \mathcal{N''} \, \Pi^-_{1234} \ket{\psi}_d$, \ie, a projection onto the subspace where $Q_{s,1}=1$. We can verify that this maps to the sparse encoding via

\begin{equation}
    \begin{split}
        Q_{s,1} \ket{\psi''} &= \ket{\psi''} \; , \\
        Q_{s,2} \ket{\psi''} &= \frac{\mathcal{N''}}{2} \left(-\gamma_5 \gamma_6 \gamma_7 \gamma_8 + \gamma_1 \cdots \gamma_8 \right)  \ket{\psi}_d   \\
        &= \frac{\mathcal{N''}}{2} \left(-\gamma_1 \gamma_2 \gamma_3 \gamma_4 + \mathbb{1}  \right) \gamma_1 \cdots \gamma_8 \ket{\psi}_d  \\
        &= \frac{\mathcal{N''}}{2} \left(-\gamma_1 \gamma_2 \gamma_3 \gamma_4 + \mathbb{1}  \right) Q_d (-i \gamma_4 \gamma_5) \ket{\psi}_d  \\
        &=\mathcal{N''} \Pi^-_{1234} \ket{\psi}_d = \ket{\psi''}   \;.
    \end{split}
\end{equation}
It remains to be shown that the information is preserved over the course of these mappings. That is, the corresponding logical basis states should be mapped to each other with no state-dependent additional amplitudes. In the sparse encoding, we can write the logical basis states as $\ket{aabb}_s$, where $a,b =\pm 1$ are the fermion parities associated with $(\text{qubit}_1, \text{ancilla}_1, \text{ancilla}_2, \text{qubit}_2)$ in that order. First, we make use of the relation $(i \gamma_3 \gamma_4) \Pi^\mp_{45} (i \gamma_3 \gamma_4) = \Pi^\pm_{45}$ in order to show

\begin{equation}
    \begin{split}
        \lVert \Pi^-_{45} \ket{aabb}_s \rVert^2 &= \bra{aabb}_s (\Pi^-_{45})^\dagger \Pi^-_{45} \ket{aabb}_s  \\
                                            &= \bra{aabb}_s  \Pi^-_{45} \ket{aabb}_s  \\
                                            &= \bra{aabb}_s a  \Pi^-_{45} a \ket{aabb}_s  \\
                                            &= \bra{aabb}_s (i\gamma_3 \gamma_4) \Pi^-_{45}  (i\gamma_3 \gamma_4) \ket{aabb}_s  \\
                                             &= \bra{aabb}_s  \Pi^+_{45}  \ket{aabb}_s  \\
                                            &= \lVert \Pi^+_{45} \ket{aabb}_s \rVert^2 \; .
    \end{split} 
\end{equation}
Together with $[i \gamma_1 \gamma_2,\Pi^\mp_{45}] = [i \gamma_7 \gamma_8,\Pi^\mp_{45}] =0$ and \eqref{eq:sparse-to-dense} this implies

\begin{equation}
    \begin{split}
    \ket{aabb}_s  &=  \Pi^-_{45} \ket{aabb}_s + \Pi^+_{45} \ket{aabb}_s   \\
                  &=  \frac{1}{\sqrt{2}} e^{i \alpha(a,b)} \ket{a \, a \oplus b \, b}_d \\
                  & \qquad \qquad \qquad + \frac{1}{\sqrt{2}} e^{i \beta(a,b)} \ket{a \, \overline{a \oplus b} \, b}_{\bar{d}}   \;,
    \end{split}
\end{equation}
where we have have denoted the dense basis states as $\ket{a \, a \oplus b \, b}_d$ with $a,b$ the parity of $\text{qubit}_1, \text{qubit}_2$. The ancilla is defined by Majoranas $\gamma_3, \gamma_6$ and has parity $a \oplus b$, where $\oplus$ denotes addition modulo 2. 
Projection with $\Pi^+_{45}$ maps into an equivalent encoding $\ket{\cdot}_{\bar{d}}$, where the total parity is odd. Furthermore, going from the sparse to the dense representation and back acts like the identity operation up to a normalization factor of 2:

\begin{figure}[h!]
    \centering
    \includegraphics[width=\columnwidth]{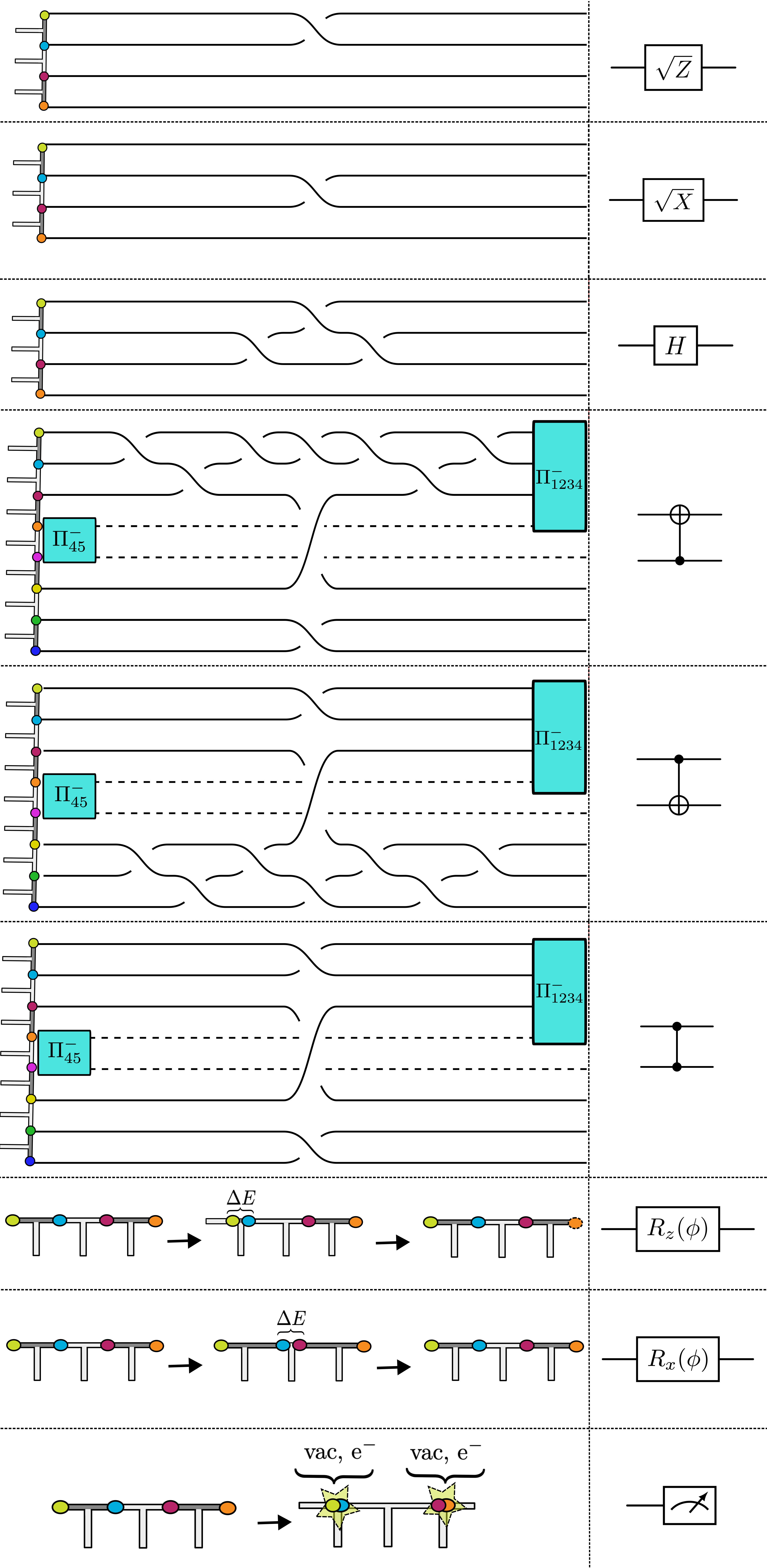}
    \caption{Glossary of braids in the even parity subspace, all implemented in the $-i\gamma_{2i-1}\gamma_{2i}$ basis as utilized in the text.
    The first three entries, correspond to a $\sqrt{Z}$, $\sqrt{X}$ and Hadamard gate respectively. 
    The next two entries correspond to controlled NOT gates, with braids enacted in the dense encoding and dashed lines, with the target being the first and second qubit respectively. 
    Subsequently, the next entry is a controlled-$Z$ gate in the dense encoding. 
    The next two entries being phase gates in the $Z$ and $X$ direction respectively, on a single sparse qubit, with the last entry being the measurement protocol, utilizing MZM fusion.}
    \label{fig:Glossary}
\end{figure}

\begin{equation}
    \begin{split}
        &\Pi^-_{1234} \Pi^-_{45}  \ket{aabb}_s  =\frac{1}{4} \big( \mathbb{1} - (i\gamma_4 \gamma_5) \\
        & +(i\gamma_1 \gamma_2)(i\gamma_3 \gamma_4)
        + (i\gamma_4 \gamma_5)(i\gamma_1 \gamma_2)(i\gamma_3 \gamma_4) \big)  \ket{aabb}_s = \\
        & \frac{1}{4} \left( \mathbb{1} - (i\gamma_4 \gamma_5) + a^2 \mathbb{1} + a^2 (i\gamma_4 \gamma_5) \right)  \ket{aabb}_s = \\
        &\frac{1}{2} \mathbb{1}  \ket{aabb}_s  \;.
    \end{split}
\end{equation}
This shows that

\begin{equation}
   \Pi^-_{1234} \ket{a \, a \oplus b \, b}_d = \frac{1}{\sqrt{2}} e^{-i \alpha(a,b)} \ket{aabb}_s  \;.
\end{equation}
Starting from any state $\ket{11bb}_s$ whose parities are $(1,1,b,b)$, we can define another basis state via $\ket{(-1)(-1)bb}_s := i \gamma_2 \gamma_3 \ket{11bb}_s$ (which corresponds to a double braid of Majoranas 2 and 3). This fixes the choice of phase. The same can be done in the dense encoding via $\ket{(-1)(-1)\oplus b b}_s := i \gamma_2 \gamma_3 \ket{1 1 \oplus b b}_s$, and analogous for the second qubit via $i \gamma_6 \gamma_7$. Crucially, $[\gamma_2 \gamma_3, \Pi^-_{45}] = [\gamma_2 \gamma_3,\Pi^-_{1234}]=0$ and, therefore, $e^{i \alpha(a,b)}= e^{i \alpha(\bar{a},b)}=e^{i \alpha(a,\bar{b})}=e^{i \alpha(\bar{a},\bar{b})}$. This ensures that if we have a sequence of braids $V$ that corresponds to a CNOT operation in the dense encoding, then $\Pi^-_{1234} V \Pi^-_{45}$ is a CNOT in the sparse encoding. 
With respect to this formalism, we provide a glossary for the set of braids to qubit gates in Fig.\,\ref{fig:Glossary}, with all two qubit gates implemented in the even parity sector of the dense encoding. 
For completeness, we also provide the relevant braids for the odd-parity sector in Fig.\,\ref{fig:Glossary-odd} with the difference between the two sectors being in the entangling controlled-$Z$ gates which are part of the CNOT gates.

\begin{figure}[t!]
    \centering
    \includegraphics[width=\columnwidth]{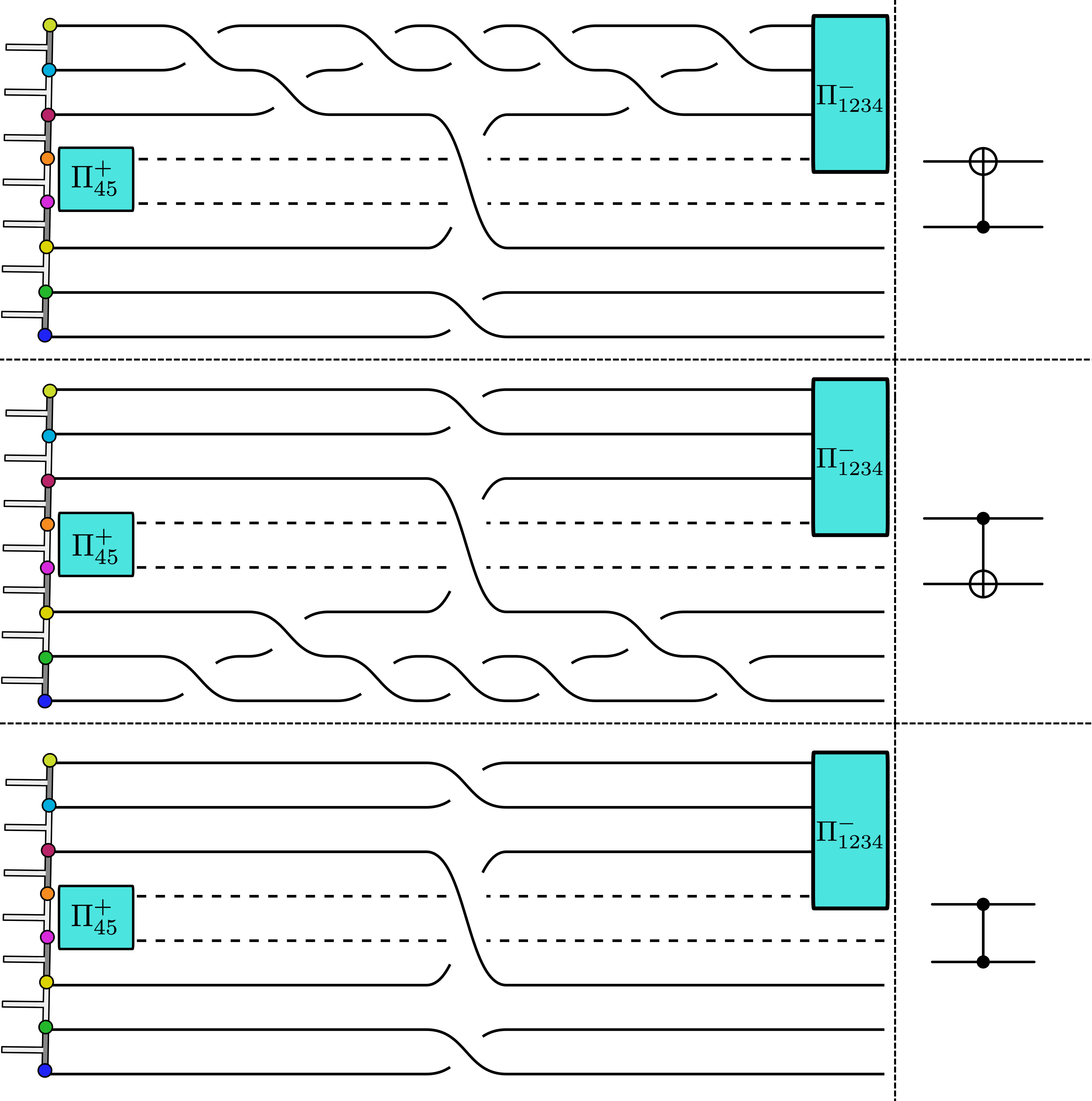}
    \caption{Glossary of braids to two-qubit entangling gates in the odd parity sector of the dense encoding.
    First two entries correspond to CNOT gates, with braids enacted in the dense encoding.  Last entry corresponds to the controlled-$Z$ gate. Braids for single-qubit gates are the same as in the even parity sector, shown in Fig.\,\ref{fig:Glossary}.}
    \label{fig:Glossary-odd}
\end{figure}


\section{Classical Simulation}\label{sec:Simulation}

Having established a general framework for mapping quantum circuits onto Majorana braiding models, we now turn to the classical simulation of such circuits. These simulations play a critical role not only in validating theoretical gate constructions but also in identifying potential sources of error, including \textit{hybridization errors}\,\cite{cheng2011,harper2019,sanno2021,bedow2023,Peeters2024,Hodge2025} and \textit{diabatic transitions}\,\cite{Brouwer2011,Scheurer2013,knapp2016,karzig-21prl057702,truong2022,Sahu25}, both of which can drive the system away from the desired quantum state. Understanding and quantifying these deviations is essential for assessing the robustness of topological gate implementations in realistic settings. These issues are examined in detail in Ref.\,\cite{Hodge2025proj}, where we simulate the dynamics of a topological quantum computer implemented on a network of Kitaev nanowires\,\cite{Kitaev2001,Alicea2012} using the methodology outlined in this section. Although the Kitaev nanowire architecture provides an excellent example, the scope of our technique extends to any system that hosts MZMs in two dimensions.

\subsection{Time-Evolution}
For a non-interacting model, we may naturally define the Hamiltonian as 
\begin{equation}
    H(t)=\frac{1}{2}\begin{pmatrix}
        c^{\dagger}_{\alpha} \quad c_{\alpha}
    \end{pmatrix}H_{\rm{BdG}}(t)
    \begin{pmatrix}
        c_{\alpha} \\ c^{\dagger}_{\alpha}
    \end{pmatrix}. \label{eq:Hamiltonian}
\end{equation}
Here, $c^{\dagger}_{\alpha}$ creates an electron with index $\alpha$, which may include degrees of freedom such as the lattice site, spin or orbital degree of freedom of the electron, with $c_{\alpha}$ the corresponding annihilation operator. 
$H_{\rm BdG}(t)$ is the matrix form for the Hamiltonian in the bare-electron basis. 
At the initial time $t=0$, there exists a diagonal basis, spanned by Bogoliubov quasiparticles, $d^{\dagger}_k$, given by
\begin{equation}
    H(0) = \sum_{k=1}^M E_k \left( d_k^\dag d_k^\pd - \frac{1}{2} \right). \label{eq:hamiltonian-diag}
\end{equation}
where $E_k\geq 0$ for all $k$. 
The connection between the bare electron basis, $\{c_\alpha\}$, and the Boguliubov quasiparticles, $\{d_k\}$, is given by the single-particle wavefunctions, $\{\psi_k\}$, which are eigenvectors of $H_{\rm BdG}(0)$, which satisfy $H_{\rm BdG}(0)\psi_k=E_k\psi_k$.
In terms of Majorana operators, we may set $d_k = (\gamma_{2k-1}+i\gamma_{2k})/2$, which is the basis stabilized by the operator $-i\gamma_{2k-1}\gamma_{2k}$, as discussed in Sec.\,\ref{sec:Sparse&Dense}. 
As such, a natural basis for the Majorana operators 
is the Fock basis of the Bogoliubov quasiparticles in Eq.\,\eqref{eq:hamiltonian-diag}.
\begin{equation}
    \ket{\vec{n}} = \prod_{k} (d_k^\dagger)^{n_k} \ket{\vec{0}_d}.
\end{equation}
After time-evolution, the state becomes
\begin{equation}
    \ket{\vec{n}(t)} = U(t, 0) \ket{\vec{n}}= \prod_{k} (d_k^\dagger(t, 0))^{n_k} \ket{\vec{0}_d(t)},
\end{equation}
where the time-evolution operator is given by
\begin{equation}
    U(t_1, t_2) = \mathcal{T} \exp \left(
        -\frac{i}{\hbar} \int_{t_2}^{t_1} dt H(t)
    \right)
\end{equation}
and the time-evolved quasiparticles are defined in an unusual convention $d_k(t,0) = U(t, 0) d_k U^\dagger(t, 0)$ since they annihilate the time-evolved quasiparticle vacuum $\ket{\vec{0}_d(t)} = U(t,0) \ket{\vec{0}_d}$.
Quasiparticles at different times are related by
\begin{eqnarray}
    \begin{pmatrix} d_i(t_1, t_2) \\ d_i^\dagger(t_1, t_2) \end{pmatrix}
    &=& \sum_{j} \begin{pmatrix}
        X_{ij} & Y_{ij}^* \\
        Y_{ij} & X_{ij}^*
    \end{pmatrix}
    \begin{pmatrix} d_j(t_3, t_4) \\ d_j^\dagger(t_3, t_4) \end{pmatrix}, \label{eq:bogo_xy} \\
    X_{ij} &=& \psi_i^\dagger(t_1, t_2) \psi_j(t_3, t_4), \\
    Y_{ij} &=& \psi_i^T(t_1, t_2) \tau_x \psi_j(t_3, t_4).
\end{eqnarray}
The single-particle wavefunctions $\psi_k(t, t_0)$ are solutions to the time-dependent Bogoliubov-de Gennes equation \cite{andreev1964,kummel1969,cheng2011},
\begin{equation}
    i\hbar \frac{\partial}{\partial t} \psi_k(t, t_0)
    = H_\text{BdG}(t) \psi_k(t, t_0),
    \label{eq:TD-BdG}
\end{equation}
where 
the initial state $\psi_k(t_0, t_0)$ are the eigenvectors of $H_\text{BdG}(t_0)$.

\subsection{Pfaffian Method}

Calculating observables essentially boils down to calculating vacuum expectation values \cite{terhal2002}.
We will consider expectation values of the form
\begin{align}
    \braket{\vec{m} | \mathcal{A} | \vec{n}(t)} = s_{\vec{m}}
    \braket{\vec{0}_d | \prod_{k} (d_k)^{m_k} \mathcal{A} \prod_{k} (d_k^\dagger)^{n_k} | \vec{0}_d(t)},
    \label{eq:vev}
\end{align}
where $s_{\vec{m}} = (-1)^{\sum_k m_k (\sum_k m_k - 1) / 2}$ is the change in sign due to reversing the order of operators in $\bra{\vec{m}}$, and $\mathcal{A}$ is an arbitrary string of fermionic operators.
Here, we stress that the vacua in Eq.\,\eqref{eq:vev} do not match.
It is convenient to use a common vacuum, which can be derived using a Bloch-Messiah decomposition \cite{bloch1962,ring1980}.
The Bogoliubov matrix in Eq.\,\eqref{eq:bogo_xy} (setting $t_1=t_2=t_4=0$, $t_3=t$) is decomposed as $X = C \bar{X} D^\dag$ and $Y = C^* \bar{Y} D^\dag$ where $C$ and $D$ are unitary and 
\begin{equation}
    \bar{X} = \begin{pmatrix}
        I & & \\
        & \oplus_k x_k \sigma_0 & \\
        & & 0
    \end{pmatrix},
    \quad
    \bar{Y} = \begin{pmatrix}
        0 & & \\
        & \oplus_k y_k i\sigma_y & \\
        & & I
    \end{pmatrix},
    \label{eq:BM}
\end{equation}
with $x_k, y_k$ positive and $x_k^2 + y_k^2 = 1$.
The three blocks are called fully empty, paired, and fully occupied, corresponding to the zero, middle, and identity blocks of $\bar{Y}$, respectively.
We define new operators
\begin{equation}
    d_i = \sum_j C_{ij} \bar{c}_j,
    \quad
    d_i(t,0) = \sum_j D_{ij} \bar{d}_j.
    \label{eq:bar-basis}
\end{equation}
We now build $\ket{\vec{0}_d(t)}$ by annihilating $\bar{d}_j$ from $\ket{\vec{0}_d}$.
\begin{equation}
    \ket{\vec{0}_d(t)} = \frac{1}{\sqrt{\mathcal{N}}} \prod_{k \in P} \bar{d}_k \bar{d}_{\bar{k}} \prod_{k \in O} \bar{d}_k \ket{\vec{0}_c},
    \label{eq:d-vacuum}
\end{equation}
where $P$ and $O$ denote paired and occupied indices, respectively, and the normalization is $\mathcal{N} = \prod_{k \in P} y_k^2$.
Here, we skip operators in the empty block since they completely annihilate the vacuum ($\bar{d}_j \ket{\vec{0}_d} = 0$ for $j$ empty).
We therefore have $d_i(t,0) \ket{\vec{0}_d(t)} = 0$ for all $i$.
The Pfaffian formula for the vacuum expectation value is given by \cite{bertsch2012,carlsson2021,Mascot2023}
\begin{equation}
\begin{split}
    \braket{\vec{m} | \mathcal{A} | \vec{n}(t)}
    &= \frac{s_{\vec{m}}}{\sqrt{\mathcal{N}(t)}} \times \\[5pt]
    &\text{pf} \begin{pmatrix}
        \wick{\c d \c d} &
        \wick{\c d \c {\mathcal{A}}} &
        \wick{\c d \c d^\dag(t)} &
        \wick{\c d \c {\bar{d}}} \\[3pt]
        &
        \wick{\c {\mathcal{A}} \c {\mathcal{A}}} &
        \wick{\c {\mathcal{A}} \c d^\dag(t)} &
        \wick{\c {\mathcal{A}} \c {\bar{d}}} \\[3pt]
        &&
        \wick{\c d^\dag(t) \c d^\dag(t)} &
        \wick{\c d^\dag(t) \c {\bar{d}}} \\[3pt]
        &&&
        \wick{\c {\bar{d}}(t) \c {\bar{d}}}
    \end{pmatrix}
\end{split}.
\end{equation}
$\wick{\c a \c b}$ denotes the matrix of contractions such that $[\wick{\c a \c b}]_{ij} = \braket{\vec{0}_d|a_i b_j|\vec{0}_d}$.
The lower triangle is obtained by anti-symmetry.

\subsection{Encoding Time-dependent Projections}
Using this architecture, we now consider the time evolution of an arbitrary Fock state $|\vec{n}\rangle$. 
Under an adiabatic time evolution, we may completely describe the time-evolved state $|\vec{n}(t)\rangle$ by considering it's connection with the initial set of many-body state space spanned by $\{|\vec{n}_k\rangle\}$ within the ground-state manifold. 
If only single-qubit braids are applied, with all unitary gates performed within the sparse encoding, all time-evolution information is contained within the transition matrix, $T(t)$, given by
\begin{equation}
    T_{\vec{m}\vec{n}}(t)=\langle \vec{m}|U(t,0)|\vec{n}\rangle=\langle \vec{m}|\vec{n}(t)\rangle.
\end{equation}
Thus, the state-space information may be read out by solving for every element of $T_{\vec{m}\vec{n}}(t)$ (where if we put constraints on the logical subspace, such as sweeping over a choice of total parity subspace, or considering transitions only within the MZM subspace, computational time may be reduced). 
This calculation may be efficiently performed via the time-dependent Pfaffian formulation introduced in the previous subsection, with previous work demonstrating this for single and two-qubit systems, using the dense encoding \cite{bedow2023, Mascot2023, Peeters2024, Crawford2024, Hodge2025}.

Suppose we consider an entangling gate, such as a CNOT or controlled-$Z$ gate. 
As discussed in \,Sec. \ref{sec:Sparse&Dense}, in the basis stabilized by $-i\gamma_{2n-1}\gamma_{2n}$, we may perform these operations on, on two qubits, by projection in the even-parity subspace, utilizing the $\Pi^-_{45}$ and $\Pi^-_{1234}$ operators defined in \eqref{eq:Projectors} 
\begin{equation}
\begin{aligned}
&\begin{aligned}
    T_{\vec{m}\vec{n}}(t)&=\bra{\vec{m}}U(t,t_2)\sqrt{2}\Pi^-_{1234}U(t_2,t_1)\sqrt{2}\Pi^-_{45}U(t_1,0)\ket{\vec{n}}\\
    &=2\bra{\vec{m}}|\Pi^-_{1234}(t,t_2)\Pi^-_{45}(t,t_1)\ket{\vec{n}(t)},
    \end{aligned}\\
&\begin{aligned}
     \Pi_{45}^{-}(t_b,t_a)&=\frac{1}{2}\left(1- i \gamma_4(t_b,t_a)\gamma_5(t_b,t_a)\right)\\
    &=d_{45}(t_b,t_a)d^{\dag}_{45}(t_b,t_a),
\end{aligned}\\
&\begin{aligned}
    \Pi_{1234}^{-}&(t_b,t_a)\\
    =&\frac{1}{2}\left(1 - \gamma_1(t_b,t_a)\gamma_2(t_b,t_a)\gamma_3(t_b,t_a)\gamma_4(t_b,t_a) \right)\\
    =&\frac{1}{2}\left(1 + Q_{s,1}(t_b,t_a) \right)
    \end{aligned}
\end{aligned}
\end{equation}
where each time-evolved operator is given by $a(t_b,t_a)=U(t_b,t_a)aU^{\dag}(t_b,t_a)$. 
Here, $U(t_1,0)$ and $U(t,t_2)$ are enacted before and after the projective measurement routine respectively, correspond to single-qubit routines performed in the sparse encoding. 
$U(t_2,t_1)$ then encodes unitary routines performed within the dense encoding. 
As shown in Fig.\,\ref{fig:Glossary}, we restrict routines performed in the dense encoding to controlled-$Z$ and CNOT gates, with all other gates performed in the sparse encoding. 
Each projection operator is scaled by a normalization factor of $\sqrt{2}$, which will conserve total probability when mapping between both encodings. 
In this way, whenever we include a sparse $\to$ dense projective measurement routine in the time evolution of the operator, the state-space information in the even parity subspace, $\{|0000\rangle,|1100\rangle, |0011\rangle, |1111\rangle\}$, may be found by considering this adjusted transition matrix. 
For example, say we incorporate $M$ two-qubit gates, sandwiched between sets of single-qubit gates, by utilizing the projective measurement routine $M$ times, throughout the quantum computation. 
The transition matrix is generalized as such: 
\begin{equation}
\begin{aligned}
    &T_{\vec{m}\vec{n}}(t)=\\
    &2^M\bra{ \vec{m}}\mathcal{T}\left(\prod^{M}_{j=1}\Pi_{1234}^{-}(t,t_{2j})\Pi_{45}^{-}(t,t_{2j-1})\right)\ket{\vec{n}(t)}\\
    &=\braket{ \vec{m}|\vec{n}_p(t)}
\end{aligned}
\end{equation}
where $\mathcal{T}$ is the time-ordering operator, defined such that 
\begin{equation}
   \mathcal{T}\big(A(t,t_i)B(t,t_j)\big)=\begin{cases}
        A(t,t_i)B(t,t_j), \quad t_i>t_j\\
        B(t,t_j)A(t,t_i), \quad t_j>t_i
    \end{cases}. 
\end{equation}
 and the state, $|\vec{n}_p(t)\rangle$, is given by $|\vec{n}_p(t)\rangle=2^M \mathcal{T}\left(\prod^{M}_{j=1}\Pi_{1234}^{-}(t,t_{2j})\Pi_{45}^{-}(t,t_{2j-1})\right)|\vec{n}(t)\rangle$. 
We will call the state $|\vec{n}_p(t)\rangle$ the \emph{projected state}, which, when included in the overlap calculation, enforces the sparse $\to$ dense encoding as part of the dynamic simulation.

We may now generalize this for an $N$ qubit system. 
The generalized form for the projection operators in \eqref{eq:Projectors}, utilized in the projective measurement routine, is given by
\begin{equation}
\begin{aligned}
    &\Pi^{\mp}_{(4i-3)(4i-2)(4i-1)4i}=\frac{1}{2}\big(1\mp\gamma_{4i-3}\gamma_{4i-2}\gamma_{4i-1}\gamma_{4i}\big),\\
    &\Pi^{\mp}_{4i(4j-3)}=\frac{1}{2}\big(1\mp i\gamma_{4i}\gamma_{4j-3}\big),
\end{aligned}
\end{equation}
where, as in the two qubit case, $\Pi^{\mp}_{(4i-3)(4i-2)(4i-1)4i}$ projects the $i$th qubit onto the even (odd) total parity subspace when indexed by $-$ ($+$). 
Similarly, $\Pi^{\mp}_{4i(4j-3)}$ projects the $i$th and $j$th qubit onto the subspace of even (odd) joint parity when indexed by $-$ ($+$).
If we set $Q_{s,i}=-\gamma_{4i-3}\gamma_{4i-2}\gamma_{4i-1}\gamma_{4i}$
we can rewrite $T_{\vec{m}\vec{n}}(t)$, after $M$ projective routines between arbitrary pairs of sparse qubits, in the condensed form: 
\begin{equation}
\begin{aligned}
    &T_{\vec{m}\vec{n}}(t)=2^M\bra{ \vec{m}}\times\\
    & 
    \begin{split}
    \mathcal{T}\big[\prod^{M}_{j=1}&\Pi_{(4\sigma^j_1-3)(4\sigma^j_1-2)(4\sigma^j_1-1)(4\sigma^j_1)}^{-}(t,t_{2j})\times\\ 
    & \Pi_{4\sigma^j_1(4\sigma^j_2-3)}^{-}(t,t_{2j-1})\big]\ket{\vec{n}(t)}
    \end{split}\\
    &=\langle \vec{m}|\mathcal{T}\big[\prod^{M}_{j=1}\sum^1_{n_j=0}\left(Q_{s,\sigma^j_1}(t,t_{2j})\right)^{n_j}\times\\
    & \quad \; d_{4\sigma^j_1(4\sigma^j_2-3)}(t,t_{2j-1})d^{\dag}_{4\sigma^j_1(4\sigma^j_2-3)}(t,t_{2j-1})\big]|\vec{n}(t)\rangle,
\end{aligned}
\end{equation}
Here, $\vec{\sigma}^j=(\sigma^j_{1},\,\sigma^j_2)$, $\sigma^j_i\in\{1,...,N\}$ is a tuple that indexes \emph{which} neighboring, sparse qubits, which we couple the $j$th time we map between the sparse and dense encoding (e.g. $\vec{\sigma}^j=(m,n)$ denotes the mapping of the $m$th, $n$th sparse qubit to the dense encoding).
This shows that the number of overlaps considered in each calculation of $T_{\vec{m}\vec{n}}(t)$ will scale as $\mathcal{O}\big(2^{M}\big)$.
In terms of classical simulation, this reveals a potential bottleneck for the method. 
For each CNOT or controlled-$Z$ gate in the routine, the computational time required to read out the final transition probability is expected to double as the number of computations also doubles in any classical simulation. 
We both test and discuss this aspect of the simulation method in \cite{Hodge2025proj}.
Here, we restrict encoding swaps to only occur when considering a string of entangling gates on two qubits, swapping back to the sparse encoding when this string is complete. 
Although the computational complexity grows exponentially with the number of entangling gates, utilizing the Pfaffian method detailed in the previous chapter scales only linearly in memory, providing a consistent and scalable method for encoding and simulating a universal gate set on an MZM based system.
\section{Discussion and outlook}\label{sec:Discussion}

In this work, we have outlined the key ingredients for universal topological quantum computation
on platforms hosting Majorana zero modes.
We reviewed the theoretical underpinnings of Majorana braiding, stabilizer encodings, and the sparse and dense logical qubit representations. 
We have emphasized the importance of transitions between these encodings, enabled via projective parity measurements, in achieving a universal gate set beyond braiding alone. To complement this semipedagogical overview, we introduced a numerically efficient simulation scheme based on Pfaffian techniques that can model the time-dependent dynamics of braiding and projective measurements. 
In \cite{Hodge2025proj}, we demonstrate the feasibility of this methodology by utilizing a Kitaev chain architecture \cite{Kitaev2001,alicea2011}, the minimal model required for braiding. 
However, we stress that this projective method is platform independent and inherently expect the same methods to be applicable on any platform able to host, manipulate, and measure MZMs. 

A central component of the sparse-to-dense encoding transition is the ability to measure the joint parity of four Majorana zero modes. Several schemes have been proposed to implement such measurements in Kitaev-like architectures. For instance, approaches based on quantum dot couplings~\cite{liu2022fusion} and dispersive readout via transmon qubits~\cite{zhuo2025readout,hassler_top-transmon_2011} allow access to multi-Majorana parity observables through energy-level shifts or controlled fusion protocols. These methods are crucial for realizing the kind of encoding transitions that enable universal computation, and their experimental feasibility is an active area of research.

While our current simulations do not explicitly model any additional architecture required to facilitate such parity measurements, our framework may be extended to do so. As long as the physical operations—including coupling, braiding, and measurement—can be represented by quadratic fermionic Hamiltonians or projectors within Gaussian fermionic states, they can in principle be incorporated into our simulation scheme. The final measurement itself is ultimately not simulatable, but measurement errors may be included stochastically in a full simulation. This opens a path toward realistic, scalable simulations of topological quantum computing protocols that include non-trivial measurement-based operations.


\begin{acknowledgements}
The authors acknowledge discussions with Bela Bauer, Lucas Hackl, Cole Peeters and Keith Wong. This work was supported by the Australian Research Council (ARC) through Grants No.\ DP200101118 and DP240100168. This research was undertaken using resources from the National Computational Infrastructure (NCI Australia), an NCRIS enabled capability supported by the Australian Government.
\end{acknowledgements}

\bibliography{bibliography/long}

\begin{thebibliography}{54}%
\makeatletter
\providecommand \@ifxundefined [1]{%
 \@ifx{#1\undefined}
}%
\providecommand \@ifnum [1]{%
 \ifnum #1\expandafter \@firstoftwo
 \else \expandafter \@secondoftwo
 \fi
}%
\providecommand \@ifx [1]{%
 \ifx #1\expandafter \@firstoftwo
 \else \expandafter \@secondoftwo
 \fi
}%
\providecommand \natexlab [1]{#1}%
\providecommand \enquote  [1]{``#1''}%
\providecommand \bibnamefont  [1]{#1}%
\providecommand \bibfnamefont [1]{#1}%
\providecommand \citenamefont [1]{#1}%
\providecommand \href@noop [0]{\@secondoftwo}%
\providecommand \href [0]{\begingroup \@sanitize@url \@href}%
\providecommand \@href[1]{\@@startlink{#1}\@@href}%
\providecommand \@@href[1]{\endgroup#1\@@endlink}%
\providecommand \@sanitize@url [0]{\catcode `\\12\catcode `\$12\catcode `\&12\catcode `\#12\catcode `\^12\catcode `\_12\catcode `\%12\relax}%
\providecommand \@@startlink[1]{}%
\providecommand \@@endlink[0]{}%
\providecommand \url  [0]{\begingroup\@sanitize@url \@url }%
\providecommand \@url [1]{\endgroup\@href {#1}{\urlprefix }}%
\providecommand \urlprefix  [0]{URL }%
\providecommand \Eprint [0]{\href }%
\providecommand \doibase [0]{https://doi.org/}%
\providecommand \selectlanguage [0]{\@gobble}%
\providecommand \bibinfo  [0]{\@secondoftwo}%
\providecommand \bibfield  [0]{\@secondoftwo}%
\providecommand \translation [1]{[#1]}%
\providecommand \BibitemOpen [0]{}%
\providecommand \bibitemStop [0]{}%
\providecommand \bibitemNoStop [0]{.\EOS\space}%
\providecommand \EOS [0]{\spacefactor3000\relax}%
\providecommand \BibitemShut  [1]{\csname bibitem#1\endcsname}%
\let\auto@bib@innerbib\@empty
\bibitem [{\citenamefont {Nayak}\ \emph {et~al.}(2008)\citenamefont {Nayak}, \citenamefont {Simon}, \citenamefont {Stern}, \citenamefont {Freedman},\ and\ \citenamefont {Sarma}}]{Nayak2008}%
  \BibitemOpen
  \bibfield  {author} {\bibinfo {author} {\bibfnamefont {C.}~\bibnamefont {Nayak}}, \bibinfo {author} {\bibfnamefont {S.~H.}\ \bibnamefont {Simon}}, \bibinfo {author} {\bibfnamefont {A.}~\bibnamefont {Stern}}, \bibinfo {author} {\bibfnamefont {M.}~\bibnamefont {Freedman}},\ and\ \bibinfo {author} {\bibfnamefont {S.~D.}\ \bibnamefont {Sarma}},\ }\bibfield  {title} {\bibinfo {title} {Non-{A}belian anyons and topological quantum computation},\ }\href {https://doi.org/10.1103/RevModPhys.80.1083} {\bibfield  {journal} {\bibinfo  {journal} {Rev. Mod. Phys.}\ }\textbf {\bibinfo {volume} {80}},\ \bibinfo {pages} {1083} (\bibinfo {year} {2008})}\BibitemShut {NoStop}%
\bibitem [{\citenamefont {Alicea}(2012)}]{Alicea2012}%
  \BibitemOpen
  \bibfield  {author} {\bibinfo {author} {\bibfnamefont {J.}~\bibnamefont {Alicea}},\ }\bibfield  {title} {\bibinfo {title} {New directions in the pursuit of {M}ajorana fermions in solid state systems},\ }\href {https://doi.org/10.1088/0034-4885/75/7/076501} {\bibfield  {journal} {\bibinfo  {journal} {Rep. Prog. Phys.}\ }\textbf {\bibinfo {volume} {75}},\ \bibinfo {pages} {076501} (\bibinfo {year} {2012})}\BibitemShut {NoStop}%
\bibitem [{\citenamefont {Aasen}\ \emph {et~al.}(2016)\citenamefont {Aasen}, \citenamefont {Hell}, \citenamefont {Mishmash}, \citenamefont {Higginbotham}, \citenamefont {Danon}, \citenamefont {Leijnse}, \citenamefont {Jespersen}, \citenamefont {Folk}, \citenamefont {Marcus}, \citenamefont {Flensberg},\ and\ \citenamefont {Alicea}}]{Aasen2016}%
  \BibitemOpen
  \bibfield  {author} {\bibinfo {author} {\bibfnamefont {D.}~\bibnamefont {Aasen}}, \bibinfo {author} {\bibfnamefont {M.}~\bibnamefont {Hell}}, \bibinfo {author} {\bibfnamefont {R.~V.}\ \bibnamefont {Mishmash}}, \bibinfo {author} {\bibfnamefont {A.}~\bibnamefont {Higginbotham}}, \bibinfo {author} {\bibfnamefont {J.}~\bibnamefont {Danon}}, \bibinfo {author} {\bibfnamefont {M.}~\bibnamefont {Leijnse}}, \bibinfo {author} {\bibfnamefont {T.~S.}\ \bibnamefont {Jespersen}}, \bibinfo {author} {\bibfnamefont {J.~A.}\ \bibnamefont {Folk}}, \bibinfo {author} {\bibfnamefont {C.~M.}\ \bibnamefont {Marcus}}, \bibinfo {author} {\bibfnamefont {K.}~\bibnamefont {Flensberg}},\ and\ \bibinfo {author} {\bibfnamefont {J.}~\bibnamefont {Alicea}},\ }\bibfield  {title} {\bibinfo {title} {Milestones toward {M}ajorana-based quantum computing},\ }\href {https://doi.org/10.1103/PhysRevX.6.031016} {\bibfield  {journal} {\bibinfo  {journal} {Phys. Rev. X}\ }\textbf {\bibinfo {volume} {6}},\ \bibinfo {pages} {031016} (\bibinfo {year}
  {2016})}\BibitemShut {NoStop}%
\bibitem [{\citenamefont {Elliott}\ and\ \citenamefont {Franz}(2015)}]{ElliottFranz2015}%
  \BibitemOpen
  \bibfield  {author} {\bibinfo {author} {\bibfnamefont {S.~R.}\ \bibnamefont {Elliott}}\ and\ \bibinfo {author} {\bibfnamefont {M.}~\bibnamefont {Franz}},\ }\bibfield  {title} {\bibinfo {title} {Colloquium: {M}ajorana fermions in nuclear, particle, and solid-state physics},\ }\href {https://doi.org/10.1103/RevModPhys.87.137} {\bibfield  {journal} {\bibinfo  {journal} {Rev. Mod. Phys.}\ }\textbf {\bibinfo {volume} {87}},\ \bibinfo {pages} {137} (\bibinfo {year} {2015})}\BibitemShut {NoStop}%
\bibitem [{\citenamefont {Beenakker}(2013)}]{Beenakker2013}%
  \BibitemOpen
  \bibfield  {author} {\bibinfo {author} {\bibfnamefont {C.~W.~J.}\ \bibnamefont {Beenakker}},\ }\bibfield  {title} {\bibinfo {title} {Search for {M}ajorana fermions in superconductors},\ }\href {https://doi.org/10.1146/annurev-conmatphys-030212-184337} {\bibfield  {journal} {\bibinfo  {journal} {Rev. Cond. Mat.}\ }\textbf {\bibinfo {volume} {4}},\ \bibinfo {pages} {113} (\bibinfo {year} {2013})}\BibitemShut {NoStop}%
\bibitem [{\citenamefont {Kitaev}(2001)}]{Kitaev2001}%
  \BibitemOpen
  \bibfield  {author} {\bibinfo {author} {\bibfnamefont {A.~Y.}\ \bibnamefont {Kitaev}},\ }\bibfield  {title} {\bibinfo {title} {Unpaired {M}ajorana fermions in quantum wires},\ }\href {https://doi.org/10.1070/1063-7869/44/10S/S29} {\bibfield  {journal} {\bibinfo  {journal} {Physics-Uspekhi}\ }\textbf {\bibinfo {volume} {44}},\ \bibinfo {pages} {131} (\bibinfo {year} {2001})}\BibitemShut {NoStop}%
\bibitem [{\citenamefont {Ivanov}(2001)}]{Ivanov2001}%
  \BibitemOpen
  \bibfield  {author} {\bibinfo {author} {\bibfnamefont {D.~A.}\ \bibnamefont {Ivanov}},\ }\bibfield  {title} {\bibinfo {title} {Non-{A}belian statistics of half-quantum vortices in p-wave superconductors},\ }\href {https://doi.org/10.1103/PhysRevLett.86.268} {\bibfield  {journal} {\bibinfo  {journal} {Phys. Rev. Lett.}\ }\textbf {\bibinfo {volume} {86}},\ \bibinfo {pages} {268} (\bibinfo {year} {2001})}\BibitemShut {NoStop}%
\bibitem [{\citenamefont {Alicea}\ \emph {et~al.}(2011)\citenamefont {Alicea}, \citenamefont {Oreg}, \citenamefont {Refael}, \citenamefont {{von Oppen}},\ and\ \citenamefont {Fisher}}]{alicea2011}%
  \BibitemOpen
  \bibfield  {author} {\bibinfo {author} {\bibfnamefont {J.}~\bibnamefont {Alicea}}, \bibinfo {author} {\bibfnamefont {Y.}~\bibnamefont {Oreg}}, \bibinfo {author} {\bibfnamefont {G.}~\bibnamefont {Refael}}, \bibinfo {author} {\bibfnamefont {F.}~\bibnamefont {{von Oppen}}},\ and\ \bibinfo {author} {\bibfnamefont {M.~P.~A.}\ \bibnamefont {Fisher}},\ }\bibfield  {title} {\bibinfo {title} {Non-{{Abelian}} statistics and topological quantum information processing in {{1D}} wire networks},\ }\href {https://doi.org/10.1038/nphys1915} {\bibfield  {journal} {\bibinfo  {journal} {Nat. Phys.}\ }\textbf {\bibinfo {volume} {7}},\ \bibinfo {pages} {412} (\bibinfo {year} {2011})}\BibitemShut {NoStop}%
\bibitem [{\citenamefont {Karzig}\ \emph {et~al.}(2017)\citenamefont {Karzig}, \citenamefont {Knapp}, \citenamefont {Lutchyn}, \citenamefont {Bonderson}, \citenamefont {Hastings}, \citenamefont {Nayak}, \citenamefont {Alicea}, \citenamefont {Flensberg}, \citenamefont {Plugge}, \citenamefont {Oreg}, \citenamefont {Marcus},\ and\ \citenamefont {Freedman}}]{Karzig2017}%
  \BibitemOpen
  \bibfield  {author} {\bibinfo {author} {\bibfnamefont {T.}~\bibnamefont {Karzig}}, \bibinfo {author} {\bibfnamefont {C.}~\bibnamefont {Knapp}}, \bibinfo {author} {\bibfnamefont {R.~M.}\ \bibnamefont {Lutchyn}}, \bibinfo {author} {\bibfnamefont {P.}~\bibnamefont {Bonderson}}, \bibinfo {author} {\bibfnamefont {M.~B.}\ \bibnamefont {Hastings}}, \bibinfo {author} {\bibfnamefont {C.}~\bibnamefont {Nayak}}, \bibinfo {author} {\bibfnamefont {J.}~\bibnamefont {Alicea}}, \bibinfo {author} {\bibfnamefont {K.}~\bibnamefont {Flensberg}}, \bibinfo {author} {\bibfnamefont {S.}~\bibnamefont {Plugge}}, \bibinfo {author} {\bibfnamefont {Y.}~\bibnamefont {Oreg}}, \bibinfo {author} {\bibfnamefont {C.~M.}\ \bibnamefont {Marcus}},\ and\ \bibinfo {author} {\bibfnamefont {M.~H.}\ \bibnamefont {Freedman}},\ }\bibfield  {title} {\bibinfo {title} {Scalable designs for quasiparticle-poisoning-protected topological quantum computation with {M}ajorana zero modes},\ }\href {https://doi.org/10.1103/PhysRevB.95.235305} {\bibfield
  {journal} {\bibinfo  {journal} {Phys. Rev. B}\ }\textbf {\bibinfo {volume} {95}},\ \bibinfo {pages} {235305} (\bibinfo {year} {2017})}\BibitemShut {NoStop}%
\bibitem [{\citenamefont {Plugge}\ \emph {et~al.}(2017)\citenamefont {Plugge}, \citenamefont {Rasmussen}, \citenamefont {Egger},\ and\ \citenamefont {Flensberg}}]{Plugge2017}%
  \BibitemOpen
  \bibfield  {author} {\bibinfo {author} {\bibfnamefont {S.}~\bibnamefont {Plugge}}, \bibinfo {author} {\bibfnamefont {A.}~\bibnamefont {Rasmussen}}, \bibinfo {author} {\bibfnamefont {R.}~\bibnamefont {Egger}},\ and\ \bibinfo {author} {\bibfnamefont {K.}~\bibnamefont {Flensberg}},\ }\bibfield  {title} {\bibinfo {title} {Majorana box qubits},\ }\href {https://doi.org/10.1088/1367-2630/aa54e1} {\bibfield  {journal} {\bibinfo  {journal} {New J. Phys.}\ }\textbf {\bibinfo {volume} {19}},\ \bibinfo {pages} {012001} (\bibinfo {year} {2017})}\BibitemShut {NoStop}%
\bibitem [{\citenamefont {Bravyi}(2006)}]{Bravyi2006}%
  \BibitemOpen
  \bibfield  {author} {\bibinfo {author} {\bibfnamefont {S.}~\bibnamefont {Bravyi}},\ }\bibfield  {title} {\bibinfo {title} {Universal quantum computation with the {$\ensuremath{\nu}=5/2$} fractional quantum {H}all state},\ }\href {https://doi.org/10.1103/PhysRevA.73.042313} {\bibfield  {journal} {\bibinfo  {journal} {Phys. Rev. A}\ }\textbf {\bibinfo {volume} {73}},\ \bibinfo {pages} {042313} (\bibinfo {year} {2006})}\BibitemShut {NoStop}%
\bibitem [{\citenamefont {Mascot}\ \emph {et~al.}(2023)\citenamefont {Mascot}, \citenamefont {Hodge}, \citenamefont {Crawford}, \citenamefont {Bedow}, \citenamefont {Morr},\ and\ \citenamefont {Rachel}}]{Mascot2023}%
  \BibitemOpen
  \bibfield  {author} {\bibinfo {author} {\bibfnamefont {E.}~\bibnamefont {Mascot}}, \bibinfo {author} {\bibfnamefont {T.}~\bibnamefont {Hodge}}, \bibinfo {author} {\bibfnamefont {D.}~\bibnamefont {Crawford}}, \bibinfo {author} {\bibfnamefont {J.}~\bibnamefont {Bedow}}, \bibinfo {author} {\bibfnamefont {D.~K.}\ \bibnamefont {Morr}},\ and\ \bibinfo {author} {\bibfnamefont {S.}~\bibnamefont {Rachel}},\ }\bibfield  {title} {\bibinfo {title} {Many-body {{Majorana}} braiding without an exponential {{Hilbert}} space},\ }\href {https://doi.org/10.1103/PhysRevLett.131.176601} {\bibfield  {journal} {\bibinfo  {journal} {Phys. Rev. Lett.}\ }\textbf {\bibinfo {volume} {131}},\ \bibinfo {pages} {176601} (\bibinfo {year} {2023})}\BibitemShut {NoStop}%
\bibitem [{\citenamefont {Hodge}\ \emph {et~al.}(2025{\natexlab{a}})\citenamefont {Hodge}, \citenamefont {Mascot}, \citenamefont {Crawford},\ and\ \citenamefont {Rachel}}]{Hodge2025}%
  \BibitemOpen
  \bibfield  {author} {\bibinfo {author} {\bibfnamefont {T.}~\bibnamefont {Hodge}}, \bibinfo {author} {\bibfnamefont {E.}~\bibnamefont {Mascot}}, \bibinfo {author} {\bibfnamefont {D.}~\bibnamefont {Crawford}},\ and\ \bibinfo {author} {\bibfnamefont {S.}~\bibnamefont {Rachel}},\ }\bibfield  {title} {\bibinfo {title} {Characterizing dynamic hybridization of {M}ajorana zero modes for universal quantum computing},\ }\href {https://doi.org/10.1103/PhysRevLett.134.096601} {\bibfield  {journal} {\bibinfo  {journal} {Phys. Rev. Lett.}\ }\textbf {\bibinfo {volume} {134}},\ \bibinfo {pages} {096601} (\bibinfo {year} {2025}{\natexlab{a}})}\BibitemShut {NoStop}%
\bibitem [{\citenamefont {Bonderson}\ \emph {et~al.}(2010)\citenamefont {Bonderson}, \citenamefont {Clarke}, \citenamefont {Nayak},\ and\ \citenamefont {Shtengel}}]{Bonderson2010}%
  \BibitemOpen
  \bibfield  {author} {\bibinfo {author} {\bibfnamefont {P.}~\bibnamefont {Bonderson}}, \bibinfo {author} {\bibfnamefont {D.~J.}\ \bibnamefont {Clarke}}, \bibinfo {author} {\bibfnamefont {C.}~\bibnamefont {Nayak}},\ and\ \bibinfo {author} {\bibfnamefont {K.}~\bibnamefont {Shtengel}},\ }\bibfield  {title} {\bibinfo {title} {Implementing arbitrary phase gates with {I}sing anyons},\ }\href {https://doi.org/10.1103/PhysRevLett.104.180505} {\bibfield  {journal} {\bibinfo  {journal} {Phys. Rev. Lett.}\ }\textbf {\bibinfo {volume} {104}},\ \bibinfo {pages} {180505} (\bibinfo {year} {2010})}\BibitemShut {NoStop}%
\bibitem [{\citenamefont {Lahtinen}\ and\ \citenamefont {Pachos}(2017)}]{lahtinen-17spp021}%
  \BibitemOpen
  \bibfield  {author} {\bibinfo {author} {\bibfnamefont {V.}~\bibnamefont {Lahtinen}}\ and\ \bibinfo {author} {\bibfnamefont {J.~K.}\ \bibnamefont {Pachos}},\ }\bibfield  {title} {\bibinfo {title} {{A Short Introduction to Topological Quantum Computation}},\ }\href {https://doi.org/10.21468/SciPostPhys.3.3.021} {\bibfield  {journal} {\bibinfo  {journal} {SciPost Phys.}\ }\textbf {\bibinfo {volume} {3}},\ \bibinfo {pages} {021} (\bibinfo {year} {2017})}\BibitemShut {NoStop}%
\bibitem [{\citenamefont {Hodge}\ \emph {et~al.}(2025{\natexlab{b}})\citenamefont {Hodge}, \citenamefont {Frey},\ and\ \citenamefont {Rachel}}]{Hodge2025proj}%
  \BibitemOpen
  \bibfield  {author} {\bibinfo {author} {\bibfnamefont {T.}~\bibnamefont {Hodge}}, \bibinfo {author} {\bibfnamefont {P.}~\bibnamefont {Frey}},\ and\ \bibinfo {author} {\bibfnamefont {S.}~\bibnamefont {Rachel}},\ }\href@noop {} {\bibinfo {title} {Projective measurements: Topological quantum computing with an arbitrary number of qubits}} (\bibinfo {year} {2025}{\natexlab{b}}),\ \bibinfo {note} {arXiv:2508.10107}\BibitemShut {NoStop}%
\bibitem [{\citenamefont {Wimmer}(2012)}]{Wimmer2012}%
  \BibitemOpen
  \bibfield  {author} {\bibinfo {author} {\bibfnamefont {M.}~\bibnamefont {Wimmer}},\ }\bibfield  {title} {\bibinfo {title} {Algorithm 923: Efficient numerical computation of the {P}faffian for dense and banded skew-symmetric matrices},\ }\href {https://doi.org/10.1145/2331130.2331138} {\bibfield  {journal} {\bibinfo  {journal} {ACM Trans. Math. Softw.}\ }\textbf {\bibinfo {volume} {38}},\ \bibinfo {pages} {30} (\bibinfo {year} {2012})}\BibitemShut {NoStop}%
\bibitem [{\citenamefont {Sarma}\ \emph {et~al.}(2015)\citenamefont {Sarma}, \citenamefont {Freedman},\ and\ \citenamefont {Nayak}}]{DasSarma2015}%
  \BibitemOpen
  \bibfield  {author} {\bibinfo {author} {\bibfnamefont {S.~D.}\ \bibnamefont {Sarma}}, \bibinfo {author} {\bibfnamefont {M.}~\bibnamefont {Freedman}},\ and\ \bibinfo {author} {\bibfnamefont {C.}~\bibnamefont {Nayak}},\ }\bibfield  {title} {\bibinfo {title} {Majorana zero modes and topological quantum computation},\ }\href {https://doi.org/10.1038/npjqi.2015.1} {\bibfield  {journal} {\bibinfo  {journal} {npj Quantum Inf}\ }\textbf {\bibinfo {volume} {1}},\ \bibinfo {pages} {15001} (\bibinfo {year} {2015})}\BibitemShut {NoStop}%
\bibitem [{\citenamefont {Field}\ and\ \citenamefont {Simula}(2018)}]{Field2018}%
  \BibitemOpen
  \bibfield  {author} {\bibinfo {author} {\bibfnamefont {B.}~\bibnamefont {Field}}\ and\ \bibinfo {author} {\bibfnamefont {T.}~\bibnamefont {Simula}},\ }\bibfield  {title} {\bibinfo {title} {Introduction to topological quantum computation with non-{A}belian anyons},\ }\href {https://doi.org/10.1088/2058-9565/aacad2} {\bibfield  {journal} {\bibinfo  {journal} {Quantum Sci. Technol.}\ }\textbf {\bibinfo {volume} {3}},\ \bibinfo {pages} {045004} (\bibinfo {year} {2018})}\BibitemShut {NoStop}%
\bibitem [{\citenamefont {Bravyi}\ and\ \citenamefont {Kitaev}(2002)}]{BravyiKitaev2002}%
  \BibitemOpen
  \bibfield  {author} {\bibinfo {author} {\bibfnamefont {S.}~\bibnamefont {Bravyi}}\ and\ \bibinfo {author} {\bibfnamefont {A.}~\bibnamefont {Kitaev}},\ }\bibfield  {title} {\bibinfo {title} {Fermionic quantum computation},\ }\href {https://doi.org/10.1006/aphy.2002.6254} {\bibfield  {journal} {\bibinfo  {journal} {Ann. Phys.}\ }\textbf {\bibinfo {volume} {298}},\ \bibinfo {pages} {210} (\bibinfo {year} {2002})}\BibitemShut {NoStop}%
\bibitem [{\citenamefont {Vijay}\ \emph {et~al.}(2015)\citenamefont {Vijay}, \citenamefont {Hsieh},\ and\ \citenamefont {Fu}}]{Vijay2015}%
  \BibitemOpen
  \bibfield  {author} {\bibinfo {author} {\bibfnamefont {S.}~\bibnamefont {Vijay}}, \bibinfo {author} {\bibfnamefont {T.~H.}\ \bibnamefont {Hsieh}},\ and\ \bibinfo {author} {\bibfnamefont {L.}~\bibnamefont {Fu}},\ }\bibfield  {title} {\bibinfo {title} {Majorana fermion surface code for universal quantum computation},\ }\href {https://doi.org/10.1103/PhysRevX.5.041038} {\bibfield  {journal} {\bibinfo  {journal} {Phys. Rev. X}\ }\textbf {\bibinfo {volume} {5}},\ \bibinfo {pages} {041038} (\bibinfo {year} {2015})}\BibitemShut {NoStop}%
\bibitem [{\citenamefont {Litinski}\ and\ \citenamefont {von Oppen}(2018)}]{Litinski2018}%
  \BibitemOpen
  \bibfield  {author} {\bibinfo {author} {\bibfnamefont {D.}~\bibnamefont {Litinski}}\ and\ \bibinfo {author} {\bibfnamefont {F.}~\bibnamefont {von Oppen}},\ }\bibfield  {title} {\bibinfo {title} {Quantum computing with {M}ajorana fermion codes},\ }\href {https://doi.org/10.1103/PhysRevB.97.205404} {\bibfield  {journal} {\bibinfo  {journal} {Phys. Rev. B}\ }\textbf {\bibinfo {volume} {97}},\ \bibinfo {pages} {205404} (\bibinfo {year} {2018})}\BibitemShut {NoStop}%
\bibitem [{\citenamefont {Gottesman}(1997)}]{Gottesman1997}%
  \BibitemOpen
  \bibfield  {author} {\bibinfo {author} {\bibfnamefont {D.}~\bibnamefont {Gottesman}},\ }\emph {\bibinfo {title} {Stabilizer Codes and Quantum Error Correction}},\ \href@noop {} {Ph.D. thesis},\ \bibinfo  {school} {California Institute of Technology} (\bibinfo {year} {1997})\BibitemShut {NoStop}%
\bibitem [{\citenamefont {Mudassar}\ \emph {et~al.}(2024)\citenamefont {Mudassar}, \citenamefont {Chien},\ and\ \citenamefont {Gottesman}}]{Mudassar2024}%
  \BibitemOpen
  \bibfield  {author} {\bibinfo {author} {\bibfnamefont {M.}~\bibnamefont {Mudassar}}, \bibinfo {author} {\bibfnamefont {R.~W.}\ \bibnamefont {Chien}},\ and\ \bibinfo {author} {\bibfnamefont {D.}~\bibnamefont {Gottesman}},\ }\bibfield  {title} {\bibinfo {title} {Encoding {M}ajorana codes},\ }\href {https://doi.org/10.1103/PhysRevA.110.032430} {\bibfield  {journal} {\bibinfo  {journal} {Phys. Rev. A}\ }\textbf {\bibinfo {volume} {110}},\ \bibinfo {pages} {032430} (\bibinfo {year} {2024})}\BibitemShut {NoStop}%
\bibitem [{\citenamefont {Aaronson}\ and\ \citenamefont {Gottesman}(2004)}]{AaronsonGottesman2004}%
  \BibitemOpen
  \bibfield  {author} {\bibinfo {author} {\bibfnamefont {S.}~\bibnamefont {Aaronson}}\ and\ \bibinfo {author} {\bibfnamefont {D.}~\bibnamefont {Gottesman}},\ }\bibfield  {title} {\bibinfo {title} {Improved simulation of stabilizer circuits},\ }\href {https://doi.org/10.1103/PhysRevA.70.052328} {\bibfield  {journal} {\bibinfo  {journal} {Phys. Rev. A}\ }\textbf {\bibinfo {volume} {70}},\ \bibinfo {pages} {052328} (\bibinfo {year} {2004})}\BibitemShut {NoStop}%
\bibitem [{\citenamefont {Gottesman}(1998)}]{Gottesman1998}%
  \BibitemOpen
  \bibfield  {author} {\bibinfo {author} {\bibfnamefont {D.}~\bibnamefont {Gottesman}},\ }\href {https://arxiv.org/abs/quant-ph/9807006} {\bibinfo {title} {The {H}eisenberg representation of quantum computers}} (\bibinfo {year} {1998}),\ \Eprint {https://arxiv.org/abs/quant-ph/9807006} {arXiv:quant-ph/9807006 [quant-ph]} \BibitemShut {NoStop}%
\bibitem [{\citenamefont {Bravyi}\ and\ \citenamefont {Kitaev}(2005)}]{BravyiKitaev2005}%
  \BibitemOpen
  \bibfield  {author} {\bibinfo {author} {\bibfnamefont {S.}~\bibnamefont {Bravyi}}\ and\ \bibinfo {author} {\bibfnamefont {A.}~\bibnamefont {Kitaev}},\ }\bibfield  {title} {\bibinfo {title} {Universal quantum computation with ideal {C}lifford gates and noisy ancillas},\ }\href {https://doi.org/10.1103/PhysRevA.71.022316} {\bibfield  {journal} {\bibinfo  {journal} {Phys. Rev. A}\ }\textbf {\bibinfo {volume} {71}},\ \bibinfo {pages} {022316} (\bibinfo {year} {2005})},\ \bibinfo {note} {\url{https://arxiv.org/abs/quant-ph/0403025}}\BibitemShut {NoStop}%
\bibitem [{\citenamefont {Bravyi}\ \emph {et~al.}(2010)\citenamefont {Bravyi}, \citenamefont {Terhal},\ and\ \citenamefont {Leemhuis}}]{Bravyi2010}%
  \BibitemOpen
  \bibfield  {author} {\bibinfo {author} {\bibfnamefont {S.}~\bibnamefont {Bravyi}}, \bibinfo {author} {\bibfnamefont {B.~M.}\ \bibnamefont {Terhal}},\ and\ \bibinfo {author} {\bibfnamefont {B.}~\bibnamefont {Leemhuis}},\ }\bibfield  {title} {\bibinfo {title} {Majorana fermion codes},\ }\href {https://doi.org/10.1088/1367-2630/12/8/083039} {\bibfield  {journal} {\bibinfo  {journal} {New J. Phys.}\ }\textbf {\bibinfo {volume} {12}},\ \bibinfo {pages} {083039} (\bibinfo {year} {2010})}\BibitemShut {NoStop}%
\bibitem [{\citenamefont {Bonderson}\ \emph {et~al.}(2008)\citenamefont {Bonderson}, \citenamefont {Freedman},\ and\ \citenamefont {Nayak}}]{Bonderson2008}%
  \BibitemOpen
  \bibfield  {author} {\bibinfo {author} {\bibfnamefont {P.}~\bibnamefont {Bonderson}}, \bibinfo {author} {\bibfnamefont {M.}~\bibnamefont {Freedman}},\ and\ \bibinfo {author} {\bibfnamefont {C.}~\bibnamefont {Nayak}},\ }\bibfield  {title} {\bibinfo {title} {Measurement-only topological quantum computation},\ }\href {https://doi.org/10.1103/PhysRevLett.101.010501} {\bibfield  {journal} {\bibinfo  {journal} {Phys. Rev. Lett.}\ }\textbf {\bibinfo {volume} {101}},\ \bibinfo {pages} {010501} (\bibinfo {year} {2008})}\BibitemShut {NoStop}%
\bibitem [{\citenamefont {Albrecht}\ \emph {et~al.}(2016)\citenamefont {Albrecht}, \citenamefont {Higginbotham}, \citenamefont {Madsen}, \citenamefont {Kuemmeth}, \citenamefont {Jespersen}, \citenamefont {Nygård}, \citenamefont {Krogstrup},\ and\ \citenamefont {Marcus}}]{Albrecht2016}%
  \BibitemOpen
  \bibfield  {author} {\bibinfo {author} {\bibfnamefont {S.~M.}\ \bibnamefont {Albrecht}}, \bibinfo {author} {\bibfnamefont {A.~P.}\ \bibnamefont {Higginbotham}}, \bibinfo {author} {\bibfnamefont {M.}~\bibnamefont {Madsen}}, \bibinfo {author} {\bibfnamefont {F.}~\bibnamefont {Kuemmeth}}, \bibinfo {author} {\bibfnamefont {T.~S.}\ \bibnamefont {Jespersen}}, \bibinfo {author} {\bibfnamefont {J.}~\bibnamefont {Nygård}}, \bibinfo {author} {\bibfnamefont {P.}~\bibnamefont {Krogstrup}},\ and\ \bibinfo {author} {\bibfnamefont {C.~M.}\ \bibnamefont {Marcus}},\ }\bibfield  {title} {\bibinfo {title} {Exponential protection of zero modes in {M}ajorana islands},\ }\href {https://doi.org/10.1038/nature17162} {\bibfield  {journal} {\bibinfo  {journal} {Nature}\ }\textbf {\bibinfo {volume} {531}},\ \bibinfo {pages} {206} (\bibinfo {year} {2016})}\BibitemShut {NoStop}%
\bibitem [{\citenamefont {Cheng}\ \emph {et~al.}(2011)\citenamefont {Cheng}, \citenamefont {Galitski},\ and\ \citenamefont {Das~Sarma}}]{cheng2011}%
  \BibitemOpen
  \bibfield  {author} {\bibinfo {author} {\bibfnamefont {M.}~\bibnamefont {Cheng}}, \bibinfo {author} {\bibfnamefont {V.}~\bibnamefont {Galitski}},\ and\ \bibinfo {author} {\bibfnamefont {S.}~\bibnamefont {Das~Sarma}},\ }\bibfield  {title} {\bibinfo {title} {Nonadiabatic effects in the braiding of non-{{Abelian}} anyons in topological superconductors},\ }\href {https://doi.org/10.1103/PhysRevB.84.104529} {\bibfield  {journal} {\bibinfo  {journal} {Phys. Rev. B}\ }\textbf {\bibinfo {volume} {84}},\ \bibinfo {pages} {104529} (\bibinfo {year} {2011})}\BibitemShut {NoStop}%
\bibitem [{\citenamefont {Zhan}\ \emph {et~al.}(2024)\citenamefont {Zhan}, \citenamefont {Mao}, \citenamefont {Chen}, \citenamefont {Yu},\ and\ \citenamefont {Luo}}]{zhan_dissipationless_2024}%
  \BibitemOpen
  \bibfield  {author} {\bibinfo {author} {\bibfnamefont {Y.-M.}\ \bibnamefont {Zhan}}, \bibinfo {author} {\bibfnamefont {G.-D.}\ \bibnamefont {Mao}}, \bibinfo {author} {\bibfnamefont {Y.-G.}\ \bibnamefont {Chen}}, \bibinfo {author} {\bibfnamefont {Y.}~\bibnamefont {Yu}},\ and\ \bibinfo {author} {\bibfnamefont {X.}~\bibnamefont {Luo}},\ }\bibfield  {title} {\bibinfo {title} {Dissipationless topological quantum computation for {Majorana} objects in sparse-dense mixed encoding process},\ }\href {https://doi.org/10.1103/PhysRevA.110.022609} {\bibfield  {journal} {\bibinfo  {journal} {Phys. Rev. A}\ }\textbf {\bibinfo {volume} {110}},\ \bibinfo {pages} {022609} (\bibinfo {year} {2024})}\BibitemShut {NoStop}%
\bibitem [{\citenamefont {Zhan}\ \emph {et~al.}(2022)\citenamefont {Zhan}, \citenamefont {Chen}, \citenamefont {Chen}, \citenamefont {Wang}, \citenamefont {Yu},\ and\ \citenamefont {Luo}}]{zhan_universal_2022}%
  \BibitemOpen
  \bibfield  {author} {\bibinfo {author} {\bibfnamefont {Y.-M.}\ \bibnamefont {Zhan}}, \bibinfo {author} {\bibfnamefont {Y.-G.}\ \bibnamefont {Chen}}, \bibinfo {author} {\bibfnamefont {B.}~\bibnamefont {Chen}}, \bibinfo {author} {\bibfnamefont {Z.}~\bibnamefont {Wang}}, \bibinfo {author} {\bibfnamefont {Y.}~\bibnamefont {Yu}},\ and\ \bibinfo {author} {\bibfnamefont {X.}~\bibnamefont {Luo}},\ }\bibfield  {title} {\bibinfo {title} {Universal topological quantum computation with strongly correlated {Majorana} edge modes},\ }\href {https://doi.org/10.1088/1367-2630/ac5f87} {\bibfield  {journal} {\bibinfo  {journal} {New J. Phys.}\ }\textbf {\bibinfo {volume} {24}},\ \bibinfo {pages} {043009} (\bibinfo {year} {2022})}\BibitemShut {NoStop}%
\bibitem [{\citenamefont {Harper}\ \emph {et~al.}(2019)\citenamefont {Harper}, \citenamefont {Pushp},\ and\ \citenamefont {Roy}}]{harper2019}%
  \BibitemOpen
  \bibfield  {author} {\bibinfo {author} {\bibfnamefont {F.}~\bibnamefont {Harper}}, \bibinfo {author} {\bibfnamefont {A.}~\bibnamefont {Pushp}},\ and\ \bibinfo {author} {\bibfnamefont {R.}~\bibnamefont {Roy}},\ }\bibfield  {title} {\bibinfo {title} {Majorana braiding in realistic nanowire {Y}-junctions and tuning forks},\ }\href {https://doi.org/10.1103/PhysRevResearch.1.033207} {\bibfield  {journal} {\bibinfo  {journal} {Phys. Rev. Res.}\ }\textbf {\bibinfo {volume} {1}},\ \bibinfo {pages} {033207} (\bibinfo {year} {2019})}\BibitemShut {NoStop}%
\bibitem [{\citenamefont {Sanno}\ \emph {et~al.}(2021)\citenamefont {Sanno}, \citenamefont {Miyazaki}, \citenamefont {Mizushima},\ and\ \citenamefont {Fujimoto}}]{sanno2021}%
  \BibitemOpen
  \bibfield  {author} {\bibinfo {author} {\bibfnamefont {T.}~\bibnamefont {Sanno}}, \bibinfo {author} {\bibfnamefont {S.}~\bibnamefont {Miyazaki}}, \bibinfo {author} {\bibfnamefont {T.}~\bibnamefont {Mizushima}},\ and\ \bibinfo {author} {\bibfnamefont {S.}~\bibnamefont {Fujimoto}},\ }\bibfield  {title} {\bibinfo {title} {Ab initio simulation of non-{A}belian braiding statistics in topological superconductors},\ }\href {https://doi.org/10.1103/PhysRevB.103.054504} {\bibfield  {journal} {\bibinfo  {journal} {Phys. Rev. B}\ }\textbf {\bibinfo {volume} {103}},\ \bibinfo {pages} {054504} (\bibinfo {year} {2021})}\BibitemShut {NoStop}%
\bibitem [{\citenamefont {Bedow}\ \emph {et~al.}(2024)\citenamefont {Bedow}, \citenamefont {Mascot}, \citenamefont {Hodge}, \citenamefont {Rachel},\ and\ \citenamefont {Morr}}]{bedow2023}%
  \BibitemOpen
  \bibfield  {author} {\bibinfo {author} {\bibfnamefont {J.}~\bibnamefont {Bedow}}, \bibinfo {author} {\bibfnamefont {E.}~\bibnamefont {Mascot}}, \bibinfo {author} {\bibfnamefont {T.}~\bibnamefont {Hodge}}, \bibinfo {author} {\bibfnamefont {S.}~\bibnamefont {Rachel}},\ and\ \bibinfo {author} {\bibfnamefont {D.~K.}\ \bibnamefont {Morr}},\ }\bibfield  {title} {\bibinfo {title} {Simulating topological quantum gates in two-dimensional magnet-superconductor hybrid structures.},\ }\href {https://doi.org/https://doi.org/10.1038/s41535-024-00703-w} {\bibfield  {journal} {\bibinfo  {journal} {npj Quantum Mater}\ }\textbf {\bibinfo {volume} {9}},\ \bibinfo {pages} {99} (\bibinfo {year} {2024})}\BibitemShut {NoStop}%
\bibitem [{\citenamefont {Peeters}\ \emph {et~al.}(2024)\citenamefont {Peeters}, \citenamefont {Hodge}, \citenamefont {Mascot},\ and\ \citenamefont {Rachel}}]{Peeters2024}%
  \BibitemOpen
  \bibfield  {author} {\bibinfo {author} {\bibfnamefont {C.}~\bibnamefont {Peeters}}, \bibinfo {author} {\bibfnamefont {T.}~\bibnamefont {Hodge}}, \bibinfo {author} {\bibfnamefont {E.}~\bibnamefont {Mascot}},\ and\ \bibinfo {author} {\bibfnamefont {S.}~\bibnamefont {Rachel}},\ }\bibfield  {title} {\bibinfo {title} {Effect of impurities and disorder on the braiding dynamics of {M}ajorana zero modes},\ }\href {https://doi.org/10.1103/PhysRevB.110.214506} {\bibfield  {journal} {\bibinfo  {journal} {Phys. Rev. B}\ }\textbf {\bibinfo {volume} {110}},\ \bibinfo {pages} {214506} (\bibinfo {year} {2024})}\BibitemShut {NoStop}%
\bibitem [{\citenamefont {Brouwer}\ \emph {et~al.}(2011)\citenamefont {Brouwer}, \citenamefont {Duckheim}, \citenamefont {Romito},\ and\ \citenamefont {von Oppen}}]{Brouwer2011}%
  \BibitemOpen
  \bibfield  {author} {\bibinfo {author} {\bibfnamefont {P.~W.}\ \bibnamefont {Brouwer}}, \bibinfo {author} {\bibfnamefont {M.}~\bibnamefont {Duckheim}}, \bibinfo {author} {\bibfnamefont {A.}~\bibnamefont {Romito}},\ and\ \bibinfo {author} {\bibfnamefont {F.}~\bibnamefont {von Oppen}},\ }\bibfield  {title} {\bibinfo {title} {Probability distribution of {M}ajorana end-state energies in disordered wires},\ }\href {https://doi.org/10.1103/PhysRevLett.107.196804} {\bibfield  {journal} {\bibinfo  {journal} {Phys. Rev. Lett.}\ }\textbf {\bibinfo {volume} {107}},\ \bibinfo {pages} {196804} (\bibinfo {year} {2011})}\BibitemShut {NoStop}%
\bibitem [{\citenamefont {Scheurer}\ and\ \citenamefont {Shnirman}(2013)}]{Scheurer2013}%
  \BibitemOpen
  \bibfield  {author} {\bibinfo {author} {\bibfnamefont {M.~S.}\ \bibnamefont {Scheurer}}\ and\ \bibinfo {author} {\bibfnamefont {A.}~\bibnamefont {Shnirman}},\ }\bibfield  {title} {\bibinfo {title} {Nonadiabatic processes in {M}ajorana qubit systems},\ }\href {https://doi.org/10.1103/PhysRevB.88.064515} {\bibfield  {journal} {\bibinfo  {journal} {Phys. Rev. B}\ }\textbf {\bibinfo {volume} {88}},\ \bibinfo {pages} {064515} (\bibinfo {year} {2013})}\BibitemShut {NoStop}%
\bibitem [{\citenamefont {Knapp}\ \emph {et~al.}(2016)\citenamefont {Knapp}, \citenamefont {Zaletel}, \citenamefont {Liu}, \citenamefont {Cheng}, \citenamefont {Bonderson},\ and\ \citenamefont {Nayak}}]{knapp2016}%
  \BibitemOpen
  \bibfield  {author} {\bibinfo {author} {\bibfnamefont {C.}~\bibnamefont {Knapp}}, \bibinfo {author} {\bibfnamefont {M.}~\bibnamefont {Zaletel}}, \bibinfo {author} {\bibfnamefont {D.~E.}\ \bibnamefont {Liu}}, \bibinfo {author} {\bibfnamefont {M.}~\bibnamefont {Cheng}}, \bibinfo {author} {\bibfnamefont {P.}~\bibnamefont {Bonderson}},\ and\ \bibinfo {author} {\bibfnamefont {C.}~\bibnamefont {Nayak}},\ }\bibfield  {title} {\bibinfo {title} {The nature and correction of diabatic errors in anyon braiding},\ }\href {https://doi.org/10.1103/PhysRevX.6.041003} {\bibfield  {journal} {\bibinfo  {journal} {Phys. Rev. X}\ }\textbf {\bibinfo {volume} {6}},\ \bibinfo {pages} {041003} (\bibinfo {year} {2016})}\BibitemShut {NoStop}%
\bibitem [{\citenamefont {Karzig}\ \emph {et~al.}(2021)\citenamefont {Karzig}, \citenamefont {Cole},\ and\ \citenamefont {Pikulin}}]{karzig-21prl057702}%
  \BibitemOpen
  \bibfield  {author} {\bibinfo {author} {\bibfnamefont {T.}~\bibnamefont {Karzig}}, \bibinfo {author} {\bibfnamefont {W.~S.}\ \bibnamefont {Cole}},\ and\ \bibinfo {author} {\bibfnamefont {D.~I.}\ \bibnamefont {Pikulin}},\ }\bibfield  {title} {\bibinfo {title} {Quasiparticle poisoning of {M}ajorana qubits},\ }\href {https://doi.org/10.1103/PhysRevLett.126.057702} {\bibfield  {journal} {\bibinfo  {journal} {Phys. Rev. Lett.}\ }\textbf {\bibinfo {volume} {126}},\ \bibinfo {pages} {057702} (\bibinfo {year} {2021})}\BibitemShut {NoStop}%
\bibitem [{\citenamefont {Truong}\ \emph {et~al.}(2023)\citenamefont {Truong}, \citenamefont {Agarwal},\ and\ \citenamefont {Pereg-Barnea}}]{truong2022}%
  \BibitemOpen
  \bibfield  {author} {\bibinfo {author} {\bibfnamefont {B.~P.}\ \bibnamefont {Truong}}, \bibinfo {author} {\bibfnamefont {K.}~\bibnamefont {Agarwal}},\ and\ \bibinfo {author} {\bibfnamefont {T.}~\bibnamefont {Pereg-Barnea}},\ }\bibfield  {title} {\bibinfo {title} {Optimizing the transport of {M}ajorana zero modes in one-dimensional topological superconductors},\ }\href {https://doi.org/10.1103/PhysRevB.107.104516} {\bibfield  {journal} {\bibinfo  {journal} {Phys. Rev. B}\ }\textbf {\bibinfo {volume} {107}},\ \bibinfo {pages} {104516} (\bibinfo {year} {2023})}\BibitemShut {NoStop}%
\bibitem [{\citenamefont {Sahu}\ and\ \citenamefont {Gangadharaiah}(2025)}]{Sahu25}%
  \BibitemOpen
  \bibfield  {author} {\bibinfo {author} {\bibfnamefont {D.}~\bibnamefont {Sahu}}\ and\ \bibinfo {author} {\bibfnamefont {S.}~\bibnamefont {Gangadharaiah}},\ }\bibfield  {title} {\bibinfo {title} {Transport of {M}ajorana bound states in the presence of telegraph noise},\ }\href {https://doi.org/10.1103/lr2b-nmrk} {\bibfield  {journal} {\bibinfo  {journal} {Phys. Rev. B}\ }\textbf {\bibinfo {volume} {111}},\ \bibinfo {pages} {235306} (\bibinfo {year} {2025})}\BibitemShut {NoStop}%
\bibitem [{\citenamefont {Andreev}(1964)}]{andreev1964}%
  \BibitemOpen
  \bibfield  {author} {\bibinfo {author} {\bibfnamefont {A.~F.}\ \bibnamefont {Andreev}},\ }\bibfield  {title} {\bibinfo {title} {Thermal conductivity of the intermediate state of superconductors},\ }\href@noop {} {\bibfield  {journal} {\bibinfo  {journal} {JETP}\ }\textbf {\bibinfo {volume} {46}},\ \bibinfo {pages} {1823} (\bibinfo {year} {1964})}\BibitemShut {NoStop}%
\bibitem [{\citenamefont {K{\"u}mmel}(1969)}]{kummel1969}%
  \BibitemOpen
  \bibfield  {author} {\bibinfo {author} {\bibfnamefont {R.}~\bibnamefont {K{\"u}mmel}},\ }\bibfield  {title} {\bibinfo {title} {Dynamics of current flow through the phase-boundary between a normal and a superconducting region},\ }\href {https://doi.org/10.1007/BF01392426} {\bibfield  {journal} {\bibinfo  {journal} {Z. Phys.}\ }\textbf {\bibinfo {volume} {218}},\ \bibinfo {pages} {472} (\bibinfo {year} {1969})}\BibitemShut {NoStop}%
\bibitem [{\citenamefont {Terhal}\ and\ \citenamefont {DiVincenzo}(2002)}]{terhal2002}%
  \BibitemOpen
  \bibfield  {author} {\bibinfo {author} {\bibfnamefont {B.~M.}\ \bibnamefont {Terhal}}\ and\ \bibinfo {author} {\bibfnamefont {D.~P.}\ \bibnamefont {DiVincenzo}},\ }\bibfield  {title} {\bibinfo {title} {Classical simulation of noninteracting-fermion quantum circuits},\ }\href {https://doi.org/10.1103/PhysRevA.65.032325} {\bibfield  {journal} {\bibinfo  {journal} {Phys. Rev. A}\ }\textbf {\bibinfo {volume} {65}},\ \bibinfo {pages} {032325} (\bibinfo {year} {2002})}\BibitemShut {NoStop}%
\bibitem [{\citenamefont {Bloch}\ and\ \citenamefont {Messiah}(1962)}]{bloch1962}%
  \BibitemOpen
  \bibfield  {author} {\bibinfo {author} {\bibfnamefont {C.}~\bibnamefont {Bloch}}\ and\ \bibinfo {author} {\bibfnamefont {A.}~\bibnamefont {Messiah}},\ }\bibfield  {title} {\bibinfo {title} {The canonical form of an antisymmetric tensor and its application to the theory of superconductivity},\ }\href {https://doi.org/10.1016/0029-5582(62)90377-2} {\bibfield  {journal} {\bibinfo  {journal} {Nucl. Phys.}\ }\textbf {\bibinfo {volume} {39}},\ \bibinfo {pages} {95} (\bibinfo {year} {1962})}\BibitemShut {NoStop}%
\bibitem [{\citenamefont {Ring}\ and\ \citenamefont {Schuck}(1980)}]{ring1980}%
  \BibitemOpen
  \bibfield  {author} {\bibinfo {author} {\bibfnamefont {P.}~\bibnamefont {Ring}}\ and\ \bibinfo {author} {\bibfnamefont {P.}~\bibnamefont {Schuck}},\ }\href@noop {} {\emph {\bibinfo {title} {The {{Nuclear Many-Body Problem}}}}},\ \bibinfo {edition} {1st}\ ed.,\ \bibinfo {number} {1864-5879}\ (\bibinfo  {publisher} {Springer Berlin, Heidelberg},\ \bibinfo {year} {1980})\BibitemShut {NoStop}%
\bibitem [{\citenamefont {Bertsch}\ and\ \citenamefont {Robledo}(2012)}]{bertsch2012}%
  \BibitemOpen
  \bibfield  {author} {\bibinfo {author} {\bibfnamefont {G.~F.}\ \bibnamefont {Bertsch}}\ and\ \bibinfo {author} {\bibfnamefont {L.~M.}\ \bibnamefont {Robledo}},\ }\bibfield  {title} {\bibinfo {title} {Symmetry restoration in {{Hartree-Fock-Bogoliubov}} based theories},\ }\href {https://doi.org/10.1103/PhysRevLett.108.042505} {\bibfield  {journal} {\bibinfo  {journal} {Phys. Rev. Lett.}\ }\textbf {\bibinfo {volume} {108}},\ \bibinfo {pages} {042505} (\bibinfo {year} {2012})}\BibitemShut {NoStop}%
\bibitem [{\citenamefont {Carlsson}\ and\ \citenamefont {Rotureau}(2021)}]{carlsson2021}%
  \BibitemOpen
  \bibfield  {author} {\bibinfo {author} {\bibfnamefont {B.~G.}\ \bibnamefont {Carlsson}}\ and\ \bibinfo {author} {\bibfnamefont {J.}~\bibnamefont {Rotureau}},\ }\bibfield  {title} {\bibinfo {title} {New and practical formulation for overlaps of {{Bogoliubov}} vacua},\ }\href {https://doi.org/10.1103/PhysRevLett.126.172501} {\bibfield  {journal} {\bibinfo  {journal} {Phys. Rev. Lett.}\ }\textbf {\bibinfo {volume} {126}},\ \bibinfo {pages} {172501} (\bibinfo {year} {2021})}\BibitemShut {NoStop}%
\bibitem [{\citenamefont {Crawford}\ \emph {et~al.}(2024)\citenamefont {Crawford}, \citenamefont {Wiesendanger},\ and\ \citenamefont {Rachel}}]{Crawford2024}%
  \BibitemOpen
  \bibfield  {author} {\bibinfo {author} {\bibfnamefont {D.}~\bibnamefont {Crawford}}, \bibinfo {author} {\bibfnamefont {R.}~\bibnamefont {Wiesendanger}},\ and\ \bibinfo {author} {\bibfnamefont {S.}~\bibnamefont {Rachel}},\ }\bibfield  {title} {\bibinfo {title} {Preparation and readout of {M}ajorana qubits in magnet-superconductor hybrid systems},\ }\href {https://doi.org/10.1103/PhysRevB.110.L220505} {\bibfield  {journal} {\bibinfo  {journal} {Phys. Rev. B}\ }\textbf {\bibinfo {volume} {110}},\ \bibinfo {pages} {L220505} (\bibinfo {year} {2024})}\BibitemShut {NoStop}%
\bibitem [{\citenamefont {Liu}\ \emph {et~al.}(2022)\citenamefont {Liu}, \citenamefont {Xiao}, \citenamefont {Kells}, \citenamefont {Katsaros},\ and\ \citenamefont {Aguado}}]{liu2022fusion}%
  \BibitemOpen
  \bibfield  {author} {\bibinfo {author} {\bibfnamefont {Y.}~\bibnamefont {Liu}}, \bibinfo {author} {\bibfnamefont {Y.}~\bibnamefont {Xiao}}, \bibinfo {author} {\bibfnamefont {G.}~\bibnamefont {Kells}}, \bibinfo {author} {\bibfnamefont {G.}~\bibnamefont {Katsaros}},\ and\ \bibinfo {author} {\bibfnamefont {R.}~\bibnamefont {Aguado}},\ }\href {https://arxiv.org/abs/2212.01653} {\bibinfo {title} {Fusion protocol for {M}ajorana modes in coupled quantum dots}} (\bibinfo {year} {2022}),\ \Eprint {https://arxiv.org/abs/2212.01653} {arXiv:2212.01653} \BibitemShut {NoStop}%
\bibitem [{\citenamefont {Zhuo}\ \emph {et~al.}(2025)\citenamefont {Zhuo}, \citenamefont {Yang}, \citenamefont {Huang}, \citenamefont {Lyu}, \citenamefont {Li}, \citenamefont {Li}, \citenamefont {Zhang}, \citenamefont {Wang}, \citenamefont {Wang}, \citenamefont {Shi} \emph {et~al.}}]{zhuo2025readout}%
  \BibitemOpen
  \bibfield  {author} {\bibinfo {author} {\bibfnamefont {E.}~\bibnamefont {Zhuo}}, \bibinfo {author} {\bibfnamefont {X.}~\bibnamefont {Yang}}, \bibinfo {author} {\bibfnamefont {Y.}~\bibnamefont {Huang}}, \bibinfo {author} {\bibfnamefont {Z.}~\bibnamefont {Lyu}}, \bibinfo {author} {\bibfnamefont {A.}~\bibnamefont {Li}}, \bibinfo {author} {\bibfnamefont {B.}~\bibnamefont {Li}}, \bibinfo {author} {\bibfnamefont {Y.}~\bibnamefont {Zhang}}, \bibinfo {author} {\bibfnamefont {X.}~\bibnamefont {Wang}}, \bibinfo {author} {\bibfnamefont {D.}~\bibnamefont {Wang}}, \bibinfo {author} {\bibfnamefont {Y.}~\bibnamefont {Shi}}, \emph {et~al.},\ }\href {https://arxiv.org/abs/2501.13367} {\bibinfo {title} {Measurement of parity-dependent energy-phase relation of the low-energy states in a potential artificial {K}itaev chain utilizing a transmon qubit}} (\bibinfo {year} {2025}),\ \Eprint {https://arxiv.org/abs/2501.13367} {arXiv:2501.13367} \BibitemShut {NoStop}%
\bibitem [{\citenamefont {Hassler}\ \emph {et~al.}(2011)\citenamefont {Hassler}, \citenamefont {Akhmerov},\ and\ \citenamefont {Beenakker}}]{hassler_top-transmon_2011}%
  \BibitemOpen
  \bibfield  {author} {\bibinfo {author} {\bibfnamefont {F.}~\bibnamefont {Hassler}}, \bibinfo {author} {\bibfnamefont {A.~R.}\ \bibnamefont {Akhmerov}},\ and\ \bibinfo {author} {\bibfnamefont {C.~W.~J.}\ \bibnamefont {Beenakker}},\ }\bibfield  {title} {\bibinfo {title} {The top-transmon: a hybrid superconducting qubit for parity-protected quantum computation},\ }\href {https://doi.org/10.1088/1367-2630/13/9/095004} {\bibfield  {journal} {\bibinfo  {journal} {New J. Phys}\ }\textbf {\bibinfo {volume} {13}},\ \bibinfo {pages} {095004} (\bibinfo {year} {2011})}\BibitemShut {NoStop}%
\end{thebibliography}%

\end{document}